\documentclass[12pt]{article}

\usepackage[natbibapa]{apacite} 

\usepackage{threeparttable}
\usepackage{epsfig, verbatim,enumerate, wrapfig}
\usepackage{graphicx}
\usepackage{epstopdf}
\usepackage{subfig}

\usepackage{tgtermes}  	% like times
\usepackage{scalerel}

\usepackage{amsmath,amssymb, amsfonts,euscript,mathrsfs, mathtools}
\usepackage{algorithm,algpseudocode}

\usepackage{pdfpages}

\usepackage{amsthm}

\usepackage{tikz-cd}

\usetikzlibrary{decorations.pathreplacing,fit,shapes.geometric}

\usepackage{todonotes}
\usepackage{mdwlist}

\usepackage[compact]{titlesec}

\titlespacing{\paragraph}{0pt}{3pt}{6pt}

\titlespacing{\section}{0pt}{15pt}{*0}
\titlespacing{\subsection}{0pt}{12pt}{*0}
\titlespacing{\subsubsection}{0pt}{12pt}{*0}

\usepackage[left=0.9in, right=0.9in,top=1in, bottom=1in]{geometry}
\usepackage{nameref,hyperref}

\usepackage{float}
\usepackage{multirow}
\floatstyle{plaintop}

\usepackage{bbm}

\usepackage{adjustbox} %%rotate tables

\usepackage{pdfpages}
\usepackage{epstopdf}

\usepackage{pifont}

\usepackage[toc,page]{appendix}

\allowdisplaybreaks

\newtheorem{assumption}{Assumption}
\title{Causal Inference using Multivariate Generalized Linear Mixed-Effects Models with Longitudinal Data
}
\author{Yizhen Xu, Jisoo Kim,  Laura K. Hummers, Ami A. Shah, Scott Zeger }

\providecommand{\keywords}[1]{\textbf{\textit{Keywords:}} #1}
\begin{document}

\maketitle
\thispagestyle{empty}
\begin{abstract}

Dynamic prediction of causal effects under different treatment regimes conditional on an individual's characteristics and longitudinal history is an essential problem in precision medicine. This is challenging in practice because outcomes and treatment assignment mechanisms are unknown in observational studies, an individual's treatment efficacy is a counterfactual, and the existence of selection bias is often unavoidable.

We propose a Bayesian framework for identifying subgroup counterfactual benefits of dynamic treatment regimes by adapting Bayesian g-computation algorithm \citep{robins1986new,zhou2019penalized} to incorporate multivariate generalized linear mixed-effects models. Unmeasured time-invariant factors are identified as subject-specific random effects in the assumed joint distribution of outcomes, time-varying confounders, and treatment assignments. Existing methods mostly assume no unmeasured confounding and  focus on balancing the observed confounder distributions between different treatments, while our method allows the presence of time-invariant unmeasured confounding. We propose a sequential ignorability assumption based on treatment assignment heterogeneity, which is analogous to balancing the latent tendency toward each treatment due to unmeasured time-invariant factors beyond the observables. We use simulation studies to assess the sensitivity of the proposed method's performance to various model assumptions. The method is applied to observational clinical data to investigate the efficacy of continuously using mycophenolate in different subgroups of scleroderma patients who were treated with the drug.
%We propose a sequential ignorability assumption conditional on the treatment assignment heterogeneity. This is analogous to balancing the distribution of observable variables as well as the latent tendency toward each treatment due to unmeasured time-invariant factors.
\end{abstract}

\keywords{Longitudinal causal inference, latent variable modeling, mixed-effects models, g-computation}

\vspace{6pt} % Needed for proper spacing betwen Keywords and first paragraph.

\section{Introduction}

% What is precision medicine, how does it connect to subgroup causal inference?     
Precision medicine \citep{kosorok2019precision,rosen2019precision} is a clinical decision-making process that uses a patient's medical history, current and previous health statuses, and observational data from a large population to make individualized treatment and care recommendations throughout the progression of a disease. For example, \citet{wang2022learning} predicted individual future biomarker trajectories and major clinical events for improving COVID-19 care, and \citet{coley2017bayesian} utilized longitudinal biomarker measurements to improve clinical decisions about whether to remove or irradiate a patient's prostate cancer. Studying the heterogeneity in an individual's treatment effect in longitudinal settings is one of the many questions of interest in precision medicine. This involves mapping patient's current information to biomarker trajectories under potential actions such as the selection and timing of therapy. We are particularly interested in using observational data to answer the causal question ``what would have happened after $\tau$ days if a specific dynamic treatment regime had been implemented, given the patient's history of $h$ days?'', where dynamic treatment regimes are defined as treatment that may change based on observed patient history. We may then determine which treatment option is the best for a patient by assessing the average treatment effect (ATE) under different regimes for a subgroup of patients who share similar characteristics or history.

Our motivating application is to study the effectiveness of an immunosuppressant medication, mycophenolate mofetil (MMF)\citep{omair2015safety,zamora2008use}, in systemic sclerosis (scleroderma) patients using clinically observed data from the Johns Hopkins Scleroderma Center Research Registry. Scleroderma is a rare multisystem autoimmune disease marked by exaggerated fibrosis, vasculopathy and derangements of the immune system. Fibrotic manifestations of the disease include skin thickening, which can compromise joint range of motion and be disabling, and interstitial lung disease (ILD), which is a major driver of morbidity and mortality. Based on limited observational and clinical trial data, MMF is the current standard of care for patients with active diffuse cutaneous systemic sclerosis and/or ILD. However, it is important to note that scleroderma is quite heterogeneous; while many patients have evidence of radiographic ILD, a relatively small percentage of patients have progressive lung disease. Given this clinical variability and the risks of MMF, such as increased risk of infection and gastrointestinal adverse effects, it remains unknown how to best deploy the treatment to the right patient at the right time. There is currently no universally accepted treatment for the disease's skin thickening due to the paucity of studies demonstrating a significant effect and the associated adverse event profiles. 

Diffuse cutaneous systemic sclerosis (dcSSc) is a subtype of scleroderma characterized by more extensive skin thickening and higher modified Rodnan skin scores, a continuous measure of skin thickening assessed in 17 body areas. In the Scleroderma Lung Study II \citep{tashkin2016mycophenolate}, which randomized patients to MMF or cyclophosphamide for scleroderma-ILD, MMF resulted in comparable improvements in forced vital capacity (FVC), a measure of lung function, and the mRSS (among dcSSc patients) by the end of 24 months. In this study, we compare the efficacy of MMF-containing versus MMF-free treatment regimens for skin and lung measurements in patients who have demonstrated tolerance to MMF, whether they have diffuse or limited/sine scleroderma. In this observational study, there are multiple practical challenges: treatment assignment is not randomized based on measured factors, biomarkers are measured irregularly, missingness patterns may be informative about biomarker values, and natural heterogeneity among subjects exists beyond what the observables can explain.
In order to tackle these issues, we use a Bayesian approach under the potential outcomes framework \citep{rubin1974estimating}, which defines causal effect as a comparison of potential outcomes for the same set of subjects under different treatment regimes. The approach has the advantages of being able to handle structural missingness, incorporating Bayesian models with the flexibility to address complex data, and naturally quantifying uncertainty, all of which are important for decision-making in precision medicine.

The primary factor in evaluating treatment efficacy, both in this and many other scenarios of comparing treatment regimes for precision medicine, is subject heterogeneity or unmeasured factors in treatment assignment and biomarker dynamics. Individual treatment decisions are intuitively sensitive to unmeasured variables that may confound disease progression. Often, the practitioner deciding on whether or not and when to treat a patient will have access to private signals about the patient's potential outcomes, such as frailty, willingness to be treated, and potential risk of adverse effects, etc. It is not always possible to assemble a set of observed variables that serve as a proxy for the available information from all of the signals. Unmeasured variables influence not only time-varying decisions but also biomarker progression.   \citet{heckman1977beta} reasoned that when unobserved permanent components exist, subjects with similar observables may have heterogenous distribution of responses, i.e. an individual's sequential responses differ systematically from the group's average behavior. %To address these challenges, it is critical to to account for patient heterogeneity in treatment assignment and disease progression for longitudinal causal inference. 

The majority of existing causal inference methods for comparing time-varying treatment assume unconfoundedness, also known as the no unmeasured confounders assumption or sequential exchangeability, i.e. the treatment assignment is independent of the potential outcomes conditional on some observed variables. The potential existence of unmeasured factors that may confound the treatment assignment and biomarker dynamics violates this fundamental assumption and thus undermines these methods, including g-estimation \citep{robins1986new,zhou2019penalized}, structural nested
models \citep{he2015structural}, history-restricted marginal structural models \citep{neugebauer2007causal}, and longitudinal targeted maximum likelihood estimation \citep{van2011targeted}. Econometric literature, on the other hand, uses unobserved effects models (UEM) or unit fixed-effects models \citep{gunasekara2014fixed,kaufman2008commentary} to eliminate time-invariant unmeasured confounding by including subject-specific intercepts and having each subject act as their own control. \citet{imai2019should} used UEM in matching to estimate contemporaneous treatment effect, i.e. comparing the outcome right before and immediately after a change in the treatment status over a short time period. The main drawback of using an UEM is that due to its assumption of strict exogeneity, it is difficult to simultaneously address biases from reverse causation and time-dependent confounding \citep{allison2017maximum}, which are common in the causal comparison of dynamic treatment regimes. 

From a modeling perspective, we account for the unmeasured patient heterogeneity in both treatment assignment and biomarker dynamics via multivariate generalized linear mixed-effects models (MGLMM) \citep{zeger1991generalized, achana2021multivariate}, which allows partial identification of unobserved permanent components through repeated measurements for each individual in a larger population. Behavioral and social science researchers have long used mixed-effects model \citep{agresti2000random,berger2004robust,luger2014robust,laird1982random,raudenbush2002hierarchical} in research involving longitudinal data . The ability of mixed-effects models to estimate subject-specific random effects allows for quantitative characterization of between-subject heterogeneity due to unobserved factors \citep{schwartz2007analysis,bolger2013intensive}.
Furthermore, these models describe the within-subject dependence in the time-varying outcome, which improves parameter estimation efficiency. However, due to the nonlinear link functions in MGLMM, estimated parameters in the generalized model often only have causal interpretations conditional on the random effects, that is, fixed-effects coefficients no longer lead to marginal causal effect based on potential outcomes\citep{greenland1999confounding} even when all covariates are exogenous \citep{zeger1988models,heagerty1999marginally}. %The fixed effect coefficients represent marginal covariate effects when the link function is either the identity link or the log link and all covariates are only associated with the response variable in an outcome model with only a random intercept \citep{gail1984biased}. 

To address this issue and enable the estimation of marginal causal effect for comparing treatment regimes with MGLMM on both population and subgroup levels, we use the g-computation algorithm, which underpins the majority of Bayesian causal inference methods. This approach directly simulates potential outcomes under a treatment path based on 
the joint distribution of time-varying confounders and outcomes conditional on patient history, consistently estimating potential outcomes and thus causal effects if all the conditional distributions are correctly specified. Standard g-estimation methods lead to biased effect estimates when unmeasured confounders are present, as the unobserved potential outcomes are not missing at random. From a sensitivity analysis perspective, \citet{yang2018sensitivity} assumes a nonidentifiable bias function quantifying the impact of unmeasured confounding on the average potential outcome under structural nested mean models. \citet{sitlani2012longitudinal} and \citet{qian2020linear} compared treatment paths that differ only at a single point in time and discussed likelihood decomposition, which supports the causal interpretation of the fixed-effects coefficients estimated from a linear mixed model, i.e. as a ``blip'' of a structural nested model. \citet{shardell2018joint} incorporated joint mixed-effects models in the g-computation algorithm to estimate the population average effect of treatment regimes over time. 

In this paper, we relax the unconfoundedness assumption and provide a framework for causal comparison of treatment paths using MGLMM, which accounts for the presence of  unmeasured time-invariant factors as latent subject heterogeneity in treatment assignments, longitudinal outcomes, and time-varying confounders. We aim to synthesize evidence from the population pertinent to clinical decisions of an individual and to account for the dynamic progression of the individual's trajectories, all while addressing the unobserved permanent factors in selection bias and adhering to the generic causal inference ideology of only using the past to infer on the current status. Existing works on causal inference with longitudinal data using mixed-effect models often marginalize over the latent components and identify causal estimand as a function of the treatment path, covariates, and fix-effects coefficients. While the unobserved stable trait factors influencing disease progression remain constant over time, our proposal dynamically updates the information relevant to these factors by sequentially estimating the subject-specific latent variables in the longitudinal outcome and time-varying confounder models based on subject's accumulating observed or counterfactual history over time. In addition, we note that the distribution of treatment assignment heterogeneity is not fully identifiable when treatment paths are binary and monotonic  because it is not a recurring process. Our discussion focuses on binary monotonic treatment process and the variance in the population distribution of treatment assignment heterogeneity is introduced as a built-in sensitivity parameter for treatment regime comparison.

Our proposal engages the treatment assignment model as part of a larger picture to bridge the gap between the confoundedness in selection bias and the heterogeneity of patients' dynamic disease progression. The work has several advantages. First, existing ways of incorporating propensity score (PS) in Bayesian causal inference \citep{li2019bayesian} include specifying outcome distribution conditional on PS \citep{zhou2019penalized}, having shared priors between propensity and outcome models, or using an inverse probability weighting or doubly robust estimator\citep{schnitzer2020data}; our method provides a new way to connect the propensity with the outcomes and time-varying confounders via the dependence structure on the subject-specific unobserved heterogeneity of the model components.  Second, our method naturally incorporates unmeasured time-invariant factors via the random effects in MGLMM, for which the estimated covariances partially inform possible existence of unmeasured confounders. Third, we provide a new perspective to investigating the impact of potential time-invariant unmeasured confounding by using the distribution of treatment assignment heterogeneity as a sensitivity parameter involved in causal estimation, rather than quantifying unmeasured confounders in post hoc sensitivity analyses \citep{robins2000sensitivity, yang2018sensitivity}. While random effects in the model components for outcomes and confounders reflect unobserved stable traits such as physiological factors of disease progression, treatment assignment heterogeneity is usually contextual and may be tractable based on knowledge about data collection and practice routine. As a result, the sensitivity parameters can be tailored to practitioners' needs as a controllable component to test the sensitivity of the causal estimates. Finally, under certain conditions, such as when treatment assignment heterogeneity is assumed to be absent and thus no unmeasured confounders exist, our approach identifies marginal subgroup treatment effect without making additional assumptions about the sensitivity parameter.

%This is enabled by the absence of orthogonality property \citep{maruo2020note} in MGLMM  due to the endogeneity of model covariates.

%Longitudinal observational studies collect repeated data on the same individuals over time for limiting bias and improving causal estimates while being representative of whole populations.

\section{Notation and Model}\label{model}

\begin{figure}[htp]
\centering
\begin{tikzcd}
Y_{ih}\arrow[rr] \arrow[rrd] \arrow[rrdd, shift right=1]  &                                                                                                                   &Y_{i,h+1} \arrow[rr] \arrow[rrd] \arrow[rrdd, shift right=1]                                                      &                                                                                                & Y_{i,h+2} \\
M_{ih}  \arrow[rr] \arrow[rru] \arrow[rrd, shift right=0.5]             &                                                                                                                   & M_{i,h+1} \arrow[rr]  \arrow[rru] \arrow[rrd, shift right=0.5]                                                                  &                                                                                                & M_{i,h+2} \\
A_{ih}\arrow[rr] \arrow[u]\arrow[uu,bend left = 60, shift right=2]  &                                                                                                                   & A_{i,h+1} \arrow[rr] \arrow[u]\arrow[uu,bend left = 60, shift right=2]                                                     &                                                                                                & A_{i,h+2} \arrow[u]\arrow[uu,bend left = 60, shift right=2]\\
                                           & b^A_i \arrow[lu, dashed] \arrow[r, no head, dashed]
                \arrow[ru, dashed]
                \arrow[rrru, dashed]  \arrow[rr, no head, dashed,  bend right] & b^M_i    \arrow[uu, dashed, bend right = 45]  \arrow[r, no head, dashed]  \arrow[dashed, rounded corners, to path={ -- ([yshift=-3.5ex]\tikztostart.south) -| ([xshift=-3.5ex]\tikztotarget.west) -- (\tikztotarget)}]{uull}
    \arrow[dashed, rounded corners, to path={ -- ([yshift=-3.5ex]\tikztostart.south) -| ([xshift=3.5ex]\tikztotarget.east) -- (\tikztotarget)}]{uurr}
    & b^Y_i    \arrow[luuu, bend right = 10, dashed] 
    \arrow[dashed, rounded corners, to path={ -- ([yshift=-5.5ex]\tikztostart.south) -| ([xshift=-5ex]\tikztotarget.west) -- (\tikztotarget)}]{uuulll}
    \arrow[dashed, rounded corners, to path={ -- ([yshift=-5.5ex]\tikztostart.south) -| ([xshift=5ex]\tikztotarget.east) -- (\tikztotarget)}]{uuur}&       
\end{tikzcd}
 \caption{Directed acyclic graph (DAG) for the generalized linear mixed model displaying temporal order of the observed variables and time-invariant unmeasured heterogeneity in both treatment assignment and biomarker dynamics. Baseline characteristics $V_i$ is excluded from the figure for simplicity.}
\label{fig:diagram}
\end{figure}
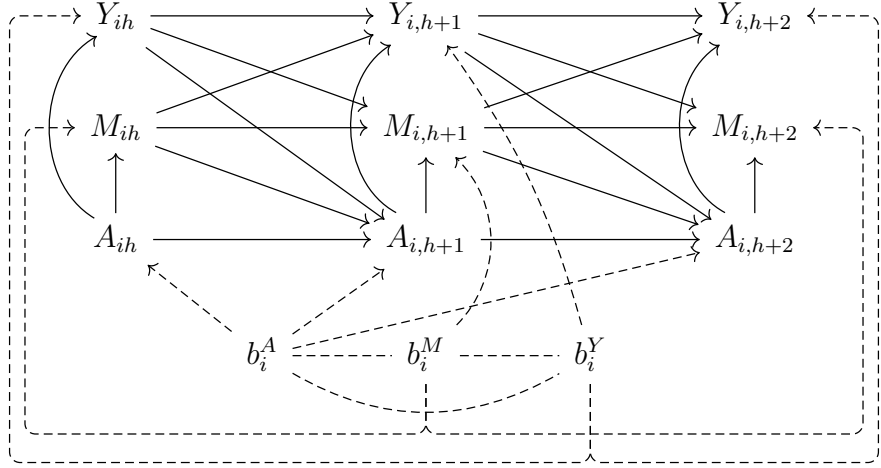 

We consider a longitudinal study that involves sequentially assigned treatment paths, and assume that time-invariant unobserved heterogeneity exists in both biomarker dynamics and treatment assignment. This paper demonstrates the method under the assumed temporal relationship of the variables as described in Figure \ref{fig:diagram}, where an arrow suggests the potential of causal relationship (single arrow) or covariance (no arrow), whereas a missing arrow implies zero influence or zero covariance. There are multivariate stochastic processes, $\{(Y_t, M_t, A_t): t\ge 0\}$, where $Y_t$, $M_t$, and $A_t$ represent the outcome process, time-dependent confounders, and sequential treatment, respectively, where $A_t\in \{0,1\}$. The  confounders are affected by previous exposure and influence future outcomes and treatment assignment. Let $\overline{Y}_{i, t_1:t_2}$, $\overline{M}_{i, t_1:t_2}$, and $\overline{A}_{i, t_1:t_2}$ denote the longitudinal paths observed for biomarkers, confounders, and interventions, respectively, during times $ t = t_1, \ldots, t_2$ for subject $i$, $i=1,\ldots, N$. At any time $t$, practitioners decide on $A_{i, (t+1)}$ based on clinical history recorded up to time $t$, i.e. past treatment path $\overline{A}_{i,0:t}$  and measurement history  $\mathcal{H}_{i,t+1}=(V_i,\overline{Y}_{i,0:t},\overline{M}_{i,0:t})$, where $V_i$ is the vector of baseline information. The updated clinical history, $(\mathcal{H}_{it},\overline{A}_{i,0:(t+1)})$, which includes the most recent treatment decision, is then the observable information for explaining the dynamics of $(Y_{i,t+1}, M_{i,t+1})$.

In this paper, we restrict the discussion to studying treatment initiations such that an initiation occurs at a single time and we assume subjects to remain treated after the initiation. Without loss of generality, we consider the outcomes to be continuous and the time-dependent confounders to be the pattern of subject visits. We model the confounders as binary variables based on the missing structure of the longitudinal outcomes. We propose using the longitudinal multivariate generalized linear mixed model (MGLMM) described below to characterize individual-level time-specific progression of biomarkers and treatment assignments. For $t=1,\ldots,T$, the continuous outcomes have a linear mixed-effects model specification,
\begin{align}\label{Ymod}
    &Y_{it} = f_A(\mathcal{H}_{it}, \overline{A}_{i,0:t}, b^Y_i; \theta^Y,\psi^Y_{it}) ,\quad \mathbb{E}(Y_{it}|\mathcal{H}_{it}, \overline{A}_{i,0:t}, b^Y_i;\theta^Y) = \lambda^{-1}_Y(\eta^Y_{it})\nonumber\\
    &\eta^Y_{it} = \phi^Y_1(\mathcal{H}_{it}) \beta^Y_1 + \phi^Y_2(\mathcal{H}_{it})\phi_A(\overline{A}_{i,0:t})^T\beta^Y_2 +  \phi^Y_3(\mathcal{H}_{it}) b^Y_{i0}+  \phi^Y_4(\mathcal{H}_{it})\phi_A(\overline{A}_{i,0:t})^T b^Y_{i1},
\end{align}
 where $\lambda_Y$ is the link function, $\phi_A(\overline{A}_{i,0:t})$ may be a function of dosage information for person $i$ at time $t$ with maximum dose $K$, e.g. $(\mathbbm{1}\{\sum^t_{s=1}A_{is}=1\}, \ldots, \mathbbm{1}\{\sum^t_{s=1}A_{is}=K\})$, $b^Y_i = (b^Y_{i0},b^Y_{i1})$ is the vector of random effects, $\psi^Y_i$ is the stochastic randomness following a mean zero distribution, e.g. $N(0,1)$, and  $\phi^Y_2(\mathcal{H}_{it})\subseteq \phi^Y_1(\mathcal{H}_{it})$, $\phi^Y_3(\mathcal{H}_{it})\subseteq \phi^Y_1(\mathcal{H}_{it})$, and $\phi^Y_4(\mathcal{H}_{it})\subseteq \phi^Y_2(\mathcal{H}_{it})$. Outcome model parameters may take the form of $\theta^Y = (\beta^Y_1,\beta^Y_2, \sigma)$, where $\sigma$ is the standard deviation of outcome distribution. 
 
 Treatment initiation is modeled as \begin{align}\label{Amod}
    &(A_{it} = 1 | A_{i,t-1} = 0)
	\sim f_A(\mathcal{H}_{it},  b^A_i; \theta^A, \psi^A_{it}), \quad \mathbb{E}(A_{it}|A_{i,t-1} = 0, \mathcal{H}_{it},  b^A_i;\theta^A) = \lambda^{-1}_A(\eta^A_{it}),\nonumber\\
	&  \eta^A_{it} =  \phi^A_1(\mathcal{H}_{it}) \beta^A_1 +  \phi^A_2(\mathcal{H}_{it}) b^A_{i0},
\end{align}
where $\lambda_A$ is the logit function, $\theta^A=(\beta^A_1,\beta^A_2)$, $b^A_i = b^A_{i0}$ is the random effect , and $\phi^A_2(\mathcal{H}_{it})\subseteq \phi^A_1(\mathcal{H}_{it})$. With a binary dependent variable, the randomness satisfies $\psi^A_{it} \sim U(0,1)$ and indicates a realization of $A_{it}$ via $ \mathbbm{1}\{\psi^A_{it} \le \lambda^{-1}_A(\eta^A_{it})\}$ under $A_{i,t-1} = 0$. We recognize that the distribution of the heterogeneity in treatment assignment, $b^A_i$, is not fully identifiable from the observed data when the assignment is not a recurrent process, i.e. happens at most once for each subject. For identifiability in model estimation, it is necessary in this instance to posit values on the variance of $b^A_i$.

For time-dependent confounders $M_{it}$, the model specification is similar to equation \eqref{Ymod},
\begin{align}\label{Mmod}
    &M_{it}
	\sim f_M(\mathcal{H}_{it}, \overline{A}_{i,0:t}, b^M_i; \theta^M, \psi^M_{it}), \quad  \mathbb{E}(M_{it}|\mathcal{H}_{it}, \overline{A}_{i,0:t}, b^M_i; \theta^M) = \lambda^{-1}_M(\eta^M_{it}) ,\nonumber\\
    &  \eta^M_{it} =  \phi^M_1(\mathcal{H}_{it}) \beta^M_1 + \phi^M_2(\mathcal{H}_{it})\phi_A(\overline{A}_{i,0:t})^T\beta^M_2 +  \phi^M_3(\mathcal{H}_{it}) b^M_{i0}+  \phi^M_4(\mathcal{H}_{it})\phi_A(\overline{A}_{i,0:t})^T b^M_{i1},
\end{align} where $\theta^M=(\beta^M_1,\beta^M_2)$ if confounders are categorical, $b^M_i = (b^M_{i0},b^M_{i1})$ is the vector of random effects, and $\phi^M_2(\mathcal{H}_{it})\subseteq \phi^M_1(\mathcal{H}_{it})$, $\phi^M_3(\mathcal{H}_{it})\subseteq \phi^M_1(\mathcal{H}_{it})$, $\phi^M_4(\mathcal{H}_{it})\subseteq \phi^M_2(\mathcal{H}_{it})$. In the motivating application, $M_{it}$ represents missing indicators of the outcomes so we set $\lambda_M$ as the logit function.
In order for binary confounders to be identifiable, $\eta^M_{it}$ has to have a parametric specification and we assume an additive model. The vector of randomness $(\psi^Y_{it},\psi^M_{it},\psi^A_{it})$ is i.i.d and independent of $b_i$; $(\psi^Y_{it},\psi^M_{it},\psi^A_{it})$ characterizes the stochasticity of counterfactual realizations. To control for stochasticity, we use the same set of randomness for causal estimation across different treatment regimes, ensuring that projected potential outcomes are comparable and reproducible.

The three model components, \eqref{Ymod}, \eqref{Amod}, and \eqref{Mmod}, are connected through a covariance structure between the random effects, 
$$b_i = (b^Y_i,b^M_i,b^A_i)^T\sim MVN(0, G_i).$$ 
In our application, we assume $G_i$ to be the same across subjects, i.e. $b_i \sim MNV(0, G)$. Subject-specific covariance $G_i$ can be realized by further parameterizing under assumed structures with individual level parameters. These random effects are interpreted as unobserved time-invariant subject-specific heterogeneity; they represent stable traits that influence the clinical trajectories and treatment assignment processes directly via random intercepts and indirectly through the effect of factors via random slopes. Specifically, $b^A_i$ is the unmeasured static heterogeneity in treatment assignment, such as a patient's frailty observed but not recorded in clinic. Without loss of generality, we assume that $\psi^Y_{it} \sim N(0,1)$, identity link for $\lambda_Y$, and logit link $\lambda_A$ and $\lambda_M$ for the rest of the manuscript.

\section{Bayesian G-Computation with MGLMM}
\subsection{Causal Quantities and Target Estimand}

 Until now, we have focused on using MGLMM to describe the data-generating mechanism as illustrated in Figure \ref{fig:diagram}. When the MGLMM is correctly specified, the posterior predictive samples of the model parameters concentrates on the true data distribution. In most cases, model parameters in MGLMM do not have a causal interpretation due to the random effects, with an exception described in supplementary material.  
 %with an exception described in supplementary material Section \ref{structural}
 %\citet{shardell2018joint} illustrated that the estimation of causal effects has to assume sequential exchangeability conditional on the random effects $b_i$ and expressed the causal quantity as a function of the parameters under certain assumptions. 

 A treatment regime dynamically defines a patient's present treatment status as a function $q(\cdot)$ of the observed or counterfactual clinical history, i.e. given a past treatment path and measurement history up to time $t$, $(\overline{A}_{i,0:(t-1)},\mathcal{H}_{it})$, the treatment sequence under regime $q$ is sequentially determined by $a_t(q) = q(\overline{A}_{i,0:(t-1)},\mathcal{H}_{it})$. For any variable $X$, $X(q)$ represents the value of $X$ had the individual received treatment under regime $q$. We define $\overline{Y}_{i,0:t}(q)$, $\overline{M}_{i,0:t}(q)$, and $\overline{a}_{0:t}(q) = (a_1(q),\ldots,a_t(q))$ as the counterfactual longitudinal trajectories of outcomes, confounders, and treatment path under regime $q$, and write the counterfactual measurement history under regime $q$ up to before time $t$ as $\mathcal{H}_{it}(q) = \{V_i, \overline{Y}_{i,t-1}(q),\overline{M}_{i,t-1}(q)\}$.

%From a modeling point of view, we recognize that the full distribution of $b^A_i$ is not identifiable from the observed data because treatment initiation happens at most once for each subject. 

The heterogeneity in treatment assignment, $b^A_i$, represents clinician-observed private signals or stable trait factors that are not captured by data but are relevant to clinicians' judgment about the potential outcomes of patients. To account for potential time-invariant unmeasured confounding that may occur naturally in the process of treating patients in clinic, we stratify the causal estimation based on treatment assignment heterogeneity. Given a specific regime of interest, $q$, and the unobserved time-invariant heterogeneity, $b^A_i$, we aim at identifying the joint distribution of a future $\tau$ days of counterfactual trajectories conditional on observed history up to a present time $h$, 
\begin{equation}\label{eq1}
P(\overline{Y}_{(h+1): (h+\tau)}(q),\overline{M}_{(h+1):(h+\tau)}(q) | V, \overline{A}_{0:h},  \overline{Y}_{0:h}, \overline{M}_{0:h}, b^A_i ).
\end{equation}
Based on \eqref{eq1} and g-computation\citep{robins1986new}, we can identify causal effects from the expectation of counterfactual outcomes 
that are functions of the fix effects parameters and the time-evolving estimations of $(b^Y_i, b^M_i)$, by integrating over observed or counterfactual histories under the treatment path determined by the regime of interest. 

Suppose we are interested in the conditional mixed average treatment effect \citep{li2022bayesian} (CMATE) within a target subgroup $T$ characterized by $\{V_i, \overline{A}_{i,0:h_i}, \overline{Y}_{i,0:h_i}, \overline{M}_{i,0:h_i}\}_{i\in T}$, in which person $i$ contributes history information up to time $h_i$ to the subgroup. For example, our application considers a subgroup $T$ that contains follow-up information prior to MMF usage from scleroderma patients who were observed to be treated with MMF. Let $\widehat{P}_T$ be the empirical distribution from the observed values in subgroup $T$, then the target quantity CMATE is defined as
\begin{align}\label{CMATE}
\int \mathbbm{E}(Y_{h_i+\tau}(q)|V_i, \overline{A}_{0:h_i}, \overline{Y}_{0:h_i}, \overline{M}_{0:h_i}) d\widehat{P}_T 
=& \frac{1}{N_T}\sum_{i\in T}\mathbbm{E}(Y_{h_i+\tau}(q)|V_i, \overline{A}_{0:h_i}, \overline{Y}_{0:h_i}, \overline{M}_{0:h_i}),
\end{align}
where $N_T$ is the number of individuals in subgroup $T$. Replacing the empirical distribution with the corresponding population distribution yields the population version of CMATE, which may be viewed as the longitudinal extension of conditional ATE and realized by jointly modeling all the variables involved in the definition of subgroup $T$. 
%$$\frac{1}{N_T}\sum_{i\in T}\int \mathbbm{E}(Y_{h_i+\tau}(q)|V_i, \overline{A}_{0:h_i}, \overline{Y}_{0:h_i}, \overline{M}_{0:h_i},b^A_i) P(b^A_i|V_i, \overline{A}_{0:h_i}, \overline{Y}_{0:h_i}, \overline{M}_{0:h_i}) db^A_i$$

% Hence, the effect conditional on random effects is not equivalent to that marginalized over the random effect distribution and accounted for in the longitudinal causal projection. However, the time-dependent covariates in the models are endogenous, i.e. influenced by previous treatment, outcomes, or confounders.We note that $b_i$ represents time-invariant heterogeneity, however, the estimation of it may change as information accumulates over time.  This is particularly challenging when we want to account for time-dependent confounders using MGLMM, because parameters in MGLMM do not imply marginal effect even when all covariates are exogenous due to the nonlinear link function \citep{zeger1988models,heagerty1999marginally}. 

\subsection{Assumptions and Method}

%We propose a procedure for estimating the causal effect given $b^A_i$ while accounting for the change of individual trajectory over time, which is essential for precision medicine. 
%can be consistently estimated without parametric form and distributional assumptions
In this section, we describe the assumptions and procedure for estimating CMATE, which jointly models multivariate time-varying components while accounting for the accumulation of individual information over time via a time-evolving update of time-invariant unobserved traits represented by $(b^Y_i, b^M_i)$. The proposal enables us to assess the sensitivity of the longitudinal causal effect estimation to different distributions of unobserved treatment heterogeneity, while allowing potential existence of time-invariant unmeasured confounding. For simplicity, we leave out subscript $i$ for the following discussion. In order to show that the conditional counterfactual joint distribution \eqref{eq1} can be identified without parametric form, we make the following assumptions:
\begin{assumption}[ ]~ %%% <-  Note that space!
For $t = 0,\ldots, T$,
\begin{enumerate}
	\item Consistency: $\overline{Y}_{0:t} = \overline{Y}_{0:t}(q)$ and $\overline{M}_{0:t} = \overline{M}_{0:t}(q)$ if $\overline{A}_{0:t} = \overline{a}_{0:t}(q)$;
	\item Positivity: $P(A_{t+1} = a_{t+1}(q) |V, \overline{A}_{0:t} = \overline{a}_{0:t}(q),\overline{Y}_{0:t}, \overline{M}_{0:t}, b^A_i) >  0$ with probability 1 for $t \ge 0$;
	\item Sequential exchangeability given $b^A_i$: for $\tau > 0$,
	\begin{align*}
	&P(\overline{Y}_{(t+1):(t+\tau)}(q), \overline{M}_{(t+1):(t+\tau)}(q) |V, A_{t+1}, \overline{A}_{0:t} = \overline{a}_{0:t}(q), \overline{Y}_{0:t}, \overline{M}_{0:t}, b^A_i)\\
	&= 	P(\overline{Y}_{(t+1):(t+\tau)}(q), \overline{M}_{(t+1):(t+\tau)}(q) |V,\overline{A}_{0:t} = \overline{a}_{0:t}(q), \overline{Y}_{0:t}, \overline{M}_{0:t}, b^A_i).
	\end{align*}
\end{enumerate}
\end{assumption}
The consistency assumption states that when the observed treatment path follows the hypothesized regime of interest, the observed and counterfactual biomarker dynamics are equivalent. It is important to note that the equivalence does not imply the same value, but rather the same distribution. Positivity guarantees that there is no systematic exclusion of a plausible treatment pattern over time. The classic assumption of sequential exchangeability \citep{greenland1986identifiability} is commonly adopted in the existing literature, assuming that the observed pretreatment history can sufficiently explain the dependence between a current treatment assignment and future counterfactuals. We extend this assumption to condition on the unobserved time-invariant heterogeneity in treatment assignment, which is quantified by the random effect $b^A_i$ in model \eqref{Amod}. In practice, the heterogeneity in treatment assignment may be attributable to patients' willingness to be treated, the potential risk of adverse effects from treatment, and the clinician's perception of treatment. Under these assumptions and that models  \eqref{Ymod}, \eqref{Amod}, and \eqref{Mmod} are correctly specified, \eqref{eq1} can be nonparametrically identified as below (see Appendix  \ref{appendix:Gformula} for details),
\begin{align}\label{gformula}
&P(\overline{Y}_{(h+1): (h+\tau)}(q),\overline{M}_{(h+1):(h+\tau)}(q) | V, \overline{A}_{0:h},  \overline{Y}_{0:h}, \overline{M}_{0:h}, b^A_i ) \nonumber\\
  =&  \prod^{h+\tau-1}_{s = h}   \int_{u_s} \int_{v_s}  P(Y_{s+1} | V, \overline{A}_{0:(s+1)} = \overline{a}_{0:(s+1)}(q), \overline{Y}_{0:s}, \overline{M}_{0:s}, b^Y_i = u_s)\nonumber\\
& \hspace{6em}  P(M_{s+1} | V, \overline{A}_{0:(s+1)} = \overline{a}_{0:(s+1)}(q), \overline{Y}_{0:s}, \overline{M}_{0:s}, b^M_i = v_s)\nonumber\\
& \hspace{6em}  P(b^Y_i = u_s, b^M_i = v_s| V,\overline{A}_{0:h}, \overline{Y}_{0:s}, \overline{M}_{0:s}, b^A_i)  du_s dv_s.
\end{align}

In Figure \ref{fig:swig}, we use a single-world intervention graph \citep{hernan2010causal,richardson2013single} (SWIG) to display the independencies that lead to \eqref{gformula} and show the counterfactual dependencies that would exist if we set the treatment path to that under regime $q$. The graph is constructed by splitting the treatment nodes $\overline{A}_{i,(h+1):(h+\tau)}$  of the causal diagram in Figure \ref{fig:diagram} and replacing all descendants of the assigned treatment with their potential outcomes, marking all counterfactuals in red. The conditional sequential exchangeability assumption is demonstrated in the SWIG by d-separation between the counterfactual trajectories $(\overline{Y}_{(h+1): (h+\tau)}(q),\overline{M}_{(h+1):(h+\tau)}(q) )$ and $A_{i,h+1}$ conditional on $(\overline{A}_{0:h}, \overline{Y}_{0:h}, \overline{M}_{0:h}, b^A_i)$. If $b^A_i$ is not controlled for, selection bias would be induced by paths $A_{i,h+1}\leftarrow b^A_i\leftrightarrow b^Y_i \rightarrow Y_{i,h+1}(q) $ and $A_{i,h+1}\leftarrow b^A_i\leftrightarrow b^M_i \rightarrow M_{i,h+1}(q) $, while stratifying on $b^A_i$ blocks these paths. Variables inside rectangles of Figure \ref{fig:swig} are quantities involved in \eqref{gformula} that are relevant to the time-evolving update of $(b^Y_i, b^M_i)$. At each time point, the distribution of $(b^Y_i, b^M_i)$ can be derived based on information backflow from observed or counterfactual biomarkers' history, resulting in a sequential update of these subject-specific unobserved permanent traits. Hypothesized treatment status $\overline{a}_{it} (q)$, $t\in [h+1,h+\tau]$, does not contribute to the sequential update of $(b^Y_i, b^M_i)$ because it generates no additional information beyond the definition of the regime of interest.

 %Moreover, fixing the treatment sequence to $\overline{a}_{0:t}(q)$ under a regime $q$ of interest may induce artificial information flow into the counterfactual biomarker dynamics through $b^A_i$.  Therefore, we create identifiability by assuming sequential exchangeability conditional on the heterogeneity in treatment assignment, $b^A_i$. 

\begin{figure}[htp]
\centering
\begin{tikzcd}[execute at end picture = {
    \node[ellipse,draw,black,inner xsep=-0.5mm, inner ysep = -2mm, xshift=-2ex,yshift=2.8ex,fit = (tikz@f@1-3-3) (tikz@f@1-4-3)] {};
    \node[ellipse,draw,black,inner xsep=2.8mm, inner ysep = -1mm,xshift=25.5ex,yshift=1.8ex,
    fit=
    (tikz@f@1-3-3) (tikz@f@1-4-3)] {}; }
                    ]
\boxed{Y_{ih}} \arrow[rr] \arrow[rrd] \arrow[rrddd, shorten = -3ex, shorten >=5ex, yshift = -5mm]                   &                                                                                                                                      & \textcolor{red}{\boxed{Y_{i,h+1}(q)}} \arrow[rr] \arrow[rrd] 
\arrow[rrddd,start anchor={[xshift=1ex]},
    end anchor={[xshift=-6.5ex,yshift=1ex]}]                                         &                             & \textcolor{red}{\boxed{Y_{i,h+2}(q)}}                                            \\
\boxed{M_{ih}} \arrow[rr] \arrow[rru] \arrow[rrdd, shorten = -2ex, shorten >=3ex, yshift = -3.5mm]                    &                                                                                                                                      & \textcolor{red}{\boxed{M_{i,h+1}(q)}} \arrow[rr] \arrow[rru] \arrow[rrdd,
    end anchor={[xshift=-5.5ex,yshift=-0.5ex]}]                                          &                             & \textcolor{red}{\boxed{M_{i,h+2}(q)}}                                            \\
\boxed{A_{ih}} \arrow[u] \arrow[uu, bend left = 60] \arrow[rrd, shorten >=1ex, yshift = -2mm] \arrow[rr,dash,
    start anchor={[xshift=16ex, yshift=-4ex]},
    end anchor={[xshift=10ex,yshift=-4ex]}
    ]&                                                                                                                                      & \textcolor{red}{a_{i,h+1}(q)} \arrow[u,shorten =1.2ex, shorten >=-0.5ex] \arrow[uu, bend left = 60] \arrow[rrd,
    end anchor={[xshift=-2.4ex,yshift=-1.5ex]}] 
    \arrow[rr,dash,
    start anchor={[xshift=15ex, yshift=-4.5ex]},
    end anchor={[xshift=12.4ex,yshift=-4.5ex]}
    ] &                             & \textcolor{red}{a_{i,h+2}(q)} \arrow[u,shorten =1ex, shorten >=-0.5ex] \arrow[uu, bend left = 60,shorten =1ex] \\
                                                              &                                                                                                                                      & A_{i,h+1}                                                                                 &                             &  {\scriptstyle A_{i,h+2}\mid a_{i,h+1}(q) }                       \\
                                                              & b^A_i \arrow[luu, dashed] \arrow[ru, dashed] \arrow[rrru, dashed] \arrow[r, no head, dashed] \arrow[rr, no head, dashed, bend right] & 
                b^M_i \arrow[r, no head, dashed] \arrow[uuu, dashed, bend right = 40]\arrow[dashed, rounded corners, to path={ -- ([yshift=-4ex]\tikztostart.south) -| ([xshift=-4ex]\tikztotarget.west) -- (\tikztotarget)}]{uuull}
    \arrow[dashed, rounded corners, to path={ -- ([yshift=-4ex]\tikztostart.south) -| ([xshift=3.5ex]\tikztotarget.east) -- (\tikztotarget)}]{uuurr}  & b^Y_i \arrow[luuuu, dashed, bend right = 15] 
    \arrow[dashed, rounded corners, to path={ -- ([yshift=-5.5ex]\tikztostart.south) -| ([xshift=-5.5ex]\tikztotarget.west) -- (\tikztotarget)}]{uuuulll}
    \arrow[dashed, rounded corners, to path={ -- ([yshift=-5.5ex]\tikztostart.south) -| ([xshift=5ex]\tikztotarget.east) -- (\tikztotarget)}]{uuuur}&
\end{tikzcd}
 \caption{SWIG.}
\label{fig:swig}
\end{figure} 

At each time $s \in [h, h+\tau)$, Monte Carlo simulation of counterfactual outcomes and confounders $(Y_{s+1}(q), M_{s+1}(q))$ based on \eqref{gformula} involves integration over $P(b^Y_i, b^M_i | V,\overline{A}_{0:h}, \overline{Y}_{0:s}, \overline{M}_{0:s}, b^A_i)$, an updated conditional posterior distribution of $(b^Y_i, b^M_i)$. The trajectories being conditioned on, $(\overline{Y}_{0:s}, \overline{M}_{0:s})$, is equivalent to $(\overline{Y}_{0:h},\overline{Y}_{(h+1):s}(q), \overline{M}_{0:h},\overline{M}_{(h+1):s}(q))$ in distribution, which is a mix of observed and counterfactual variables. Note that the counterfactual trajectories $(\overline{Y}_{(h+1):s}(q), \overline{M}_{(h+1):s}(q))$  have the following distribution 
$$  \prod^{s-1}_{s = h} P( Y_{s+1} , M_{s+1}| V, \overline{A}_{0:(s+1)} = \overline{a}_{0:(s+1)}(q), \overline{Y}_{0:s}, \overline{M}_{0:s}, b^A_i). $$
under regime $q$ and treatment assignment heterogeneity $b^A_i$, as implied by the formulation of equation \eqref{gformula}. The sampling of $(b^Y_i, b^M_i)\sim P(b^Y_i, b^M_i | V,\overline{A}_{0:h}, \overline{Y}_{0:s}, \overline{M}_{0:s}, b^A_i)$ may be complicated by nonlinear link functions in the MGLMM. We consider the following general strategy: first, calculate the Laplace approximation of the posterior distribution $(b^Y_i, b^M_i, b^A_i| V,\overline{A}_{0:h}, \overline{Y}_{0:s}, \overline{M}_{0:s})$, denoted by $MVN(\hat{b}_i, V)$, and then sample the heterogeneities via the corresponding conditional distribution, $(b^Y_i, b^M_i)|b^A_i$, with $b^A_i$ set to a certain value. The procedure is illustrated in Appendix \ref{appendix:sequential}. We provide in Appendix \ref{appendix:pseudo} the pseudocode for generating posterior samples of counterfactual trajectories from $P(\overline{Y}_{(h+1): (h+\tau)}(q),\overline{M}_{(h+1):(h+\tau)}(q) | V, \overline{A}_{0:h},  \overline{Y}_{0:h}, \overline{M}_{0:h}, b^A_i)$ based on \eqref{gformula}.

We have been stratifying on $b^A_i$ thus far in our discussion. The target estimand CMATE expressed in equation \eqref{CMATE} is the marginal subgroup ATE, marginalizing over unobserved heterogeneity. Therefore, using the following formula, we integrate each component of CMATE over the distribution of $b^A_i$ conditional on subgroup $T$,
\begin{align}\label{int}
   &\mathbbm{E}(Y_{h+\tau}(q)|V, \overline{A}_{0:h}, \overline{Y}_{0:h}, \overline{M}_{0:h})\nonumber\\
 =&\int_{w} \mathbbm{E}(Y_{h+\tau}(q)|V, \overline{A}_{0:h}, \overline{Y}_{0:h}, \overline{M}_{0:h}, b^A_i=w)P(b^A_i=w|V, \overline{A}_{0:h}, \overline{Y}_{0:h}, \overline{M}_{0:h}) dw.
\end{align}
As a result, CMATE is a functional of the counterfactual joint distribution \eqref{eq1} because the conditional expectation $\mathbbm{E}(Y_{h+\tau}(q)|V, \overline{A}_{0:h}, \overline{Y}_{0:h}, \overline{M}_{0:h}, b^A_i)$ in \eqref{int} can be expressed as
\begin{align*}
   & \mathbbm{E}(Y_{h+\tau}(q)|V, \overline{A}_{0:h}, \overline{Y}_{0:h}, \overline{M}_{0:h}, b^A_i)\\
   = & \int_{y_{h+\tau}}\int_{m_{h+\tau}}\ldots  \int_{y_{h+1}}\int_{m_{h+1}} \\
   & \hspace{1em}y_{h+\tau} P(\overline{Y}_{(h+1):(h+\tau)}(q) = \overline{y}_{(h+1):(h+\tau)}, \overline{M}_{(h+1):(h+\tau)}(q) = \overline{m}_{(h+1):(h+\tau)} | V, \overline{A}_{0:h}, \overline{Y}_{0:h}, \overline{M}_{0:h}, b^A_i)\\
   &\hspace{25em}  dm_{h+1} dy_{h+1}\ldots d m_{h+\tau} d y_{h+\tau}.
\end{align*}
When comparing regimes $q_1$ and $q_2$, we estimate the causal contrast by integrating $$\mathbbm{E}(Y_{h+\tau}(q_1)|V, \overline{A}_{0:h}, \overline{Y}_{0:h}, \overline{M}_{0:h}, b^A_i) - \mathbbm{E}(Y_{h+\tau}(q_2)|V, \overline{A}_{0:h}, \overline{Y}_{0:h}, \overline{M}_{0:h}, b^A_i)$$%given history information up to a time $h$, baseline characteristics, and the time-invariant assignment heterogeneity $b^A_i$.  
over the subgroup distribution of treatment assignment heterogeneity, $P(b^A_i|V, \overline{A}_{0:h}, \overline{Y}_{0:h}, \overline{M}_{0:h}) $. Appendix \ref{appendix:pseudo} contains the computational details for calculating CMATE, and the motivating application illustrates subgroup causal effect estimation under the proposed method.

The subgroup distribution of $b^A_i$, which is controlled for and marginalized over, serves as the sensitivity parameter in the calculation of CMATE. Recall that MGLMM assumes that $(b^Y_i,b^M_i,b^A_i)$ follows $MVN(0,G)$. Let $v$ denote the variance of $b^A_i$, representing the assumed amount of variation in treatment assignment heterogeneity among subjects. When treatment assignment is a binary monotonic process as in the motivating application, $v$ is unidentifiable and needs a posited value in model estimation because treatment assignment is not a recurring event. When subgroup $T$ contains individual history of different lengths,  subgroup distribution $P(b^A_i|V, \overline{A}_{0:h}, \overline{Y}_{0:h}, \overline{M}_{0:h})$ can be derived conditional on a given value of $v$ for each individual. The evaluation of causal effectiveness may vary under different values of $v$ . In this case, the variance of the unidentifiable time-invariant quantity $b^A_i$ serves as a sensitivity parameter in the estimation of causal effects. For population ATE, the target quantity can be derived as
\begin{align*}
   \mathbbm{E}(Y_{h+\tau}(q))
 =&  \int_{w} \mathbbm{E}(Y_{h+\tau}(q)|, b^A_i=w)P(b^A_i=w) dw ,
\end{align*}
where $b^A_i\sim N(0,v)$ based on model assumption. Appendix \ref{appendix:Gformula} gives further details to the calculation of the mixed ATE of the target population \citep{li2022bayesian}, $\widehat{\mathbbm{E}}(Y_{h+\tau}(q))$, which replaces the target population distribution with the corresponding empirical distribution.

%Based on model specification, the marginal population distribution of treatment assignment heterogeneity, $b^A_i$, is Gaussian distributed with variance $v$. Thus, the population ATE $\mathbbm{E}(Y_{h+\tau}(q))$ can be obtained by integrating \eqref{eq1} over subgroup variables $( V, \overline{A}_{0:h},  \overline{Y}_{0:h}, \overline{M}_{0:h})$, counterfactual history  $(\overline{Y}_{(h+1):(h+\tau-1)}(q), \overline{M}_{(h+1):(h+\tau-1)}(q))$, and treatment assignment heterogeneity $b^A_i \sim N(0,v)$. 

%Specifying $v$ creates identifiability for conditional subgroup ATE such as $\mathbbm{E}(Y_{h+\tau}(q)|V, \overline{A}_{0:h}, \overline{Y}_{0:h}, \overline{M}_{0:h}, b^A_i)$ and marginal population ATE $\mathbbm{E}(Y_{h+\tau}(q))$. Given the parametric assumption of MGLMM and $v$, distribution $P(b^A_i|V, \overline{A}_{0:h}, \overline{Y}_{0:h}, \overline{M}_{0:h})$ is estimable conditional on history information for the calculation of the marginal subgroup ATE

When no treatment heterogeneity under MGLMM, i.e. setting $v=0$, the proposal simplifies to the standard g-computation of utilizing only the model components for outcomes and confounders because the assignment mechanism is unconfounded \citep{li2022bayesian}. Having $v=0$ is a sufficient but unnecessary condition for having no unmeasured confounders. Under MGLMM, $\text{cov}(b^A_i, b^M_i) =\text{cov}(b^A_i, b^Y_i) = 0$ leads to no unmeasured confounders. When there are no unmeasured confounders, MGLMM still allows unobserved factors to influence treatment assignment, i.e. $v\neq 0$, as long as $b^A_i$ is not correlated with the unobserved heterogeneity in biomarker dynamics $(b^Y_i, b^M_i)$; examples of such non-confounding treatment assignment heterogeneity include preference for a treatment based on personal beliefs or social stigma. On the other hand, we note that the covariances $\text{cov}(b^A_i, b^M_i)$ and $\text{cov}(b^A_i, b^Y_i)$ are estimable given the variance of $b^A_i$, $v$. As a result, our method is able to provide insight into the potential existence of unmeasured confounders based on the MGLMM's estimated covariances.

\section{Simulation}

Assuming each person has two follow-up visits, $T_i=2$, we simulate continuous biomarker $Y_{it}$ and binary time-varying treatment $A_{it}$ via
\begin{align*}
&Y_{it} = 0.4- 0.3V_i - 0.1 t +  \sum^2_{k=1} \frac{\nu_k}{2} \times \mathbbm{1}\big\{ \sum^t_{s=1}A_{is}=k\big\}+0.4Y_{i,t-1}+ b^Y_{i0} +  e^Y_{it} \text{  and}\\
& \text{logit}\{P(A_{it}(s) = 1 |A_{i,t-1}(s)=0 )\} = -0.1V_i 	- 0.5 t -0.35 Y_{i,t-1} +b^A_{i0},
\end{align*}
where $e^Y_{it} \sim N(0,0.4^2)$, $V_i$ is the baseline covariate, $V_i \sim \text{Bernoulli}(0.5)$, and $Y_{i0} \sim N(0,1)$.  Write $\rho = \text{Corr}(b^A_{i0}, b^Y_{i0})$, $s_A = \sqrt{\text{Var}(b^A_{i0})}$, $s_Y = \sqrt{\text{Var}(b^Y_{i0})}$. We assume the random effects to follow $(b^A_{i0}, b^Y_{i0}) \sim N(0, G)$, where the covariance matrix $G$ has elements $G_{11} = s^2_A, G_{12} = G_{21} = \rho s_A s_Y$, and $G_{22}=s^2_Y$. We set $s_Y = 0.8$ and randomly sample 100 replicates for each of the 101 settings defined by parameter combinations $(s_A, \rho) \in (0,0) \cup \{(s_A, \rho); s_A \in \{0.1,\ldots,0.9,1\} , \rho\in \{0,0.1,\ldots,0.9\}\}$. When $s_A>0$ and $\rho < 1$,  $G$ is guaranteed to be positive definite because $\text{det}(G) = (1-\rho)s^2_A > 0$. When $\rho=0$, matrix $G = \begin{pmatrix} 
s^2_A &0 \\ 
 0& s^2_Y
\end{pmatrix}$ represents the case of no unmeasured confounding. We consider two scenarios, $\nu_k=k$ and $\nu_k=0$ for $k = 1,2$, where $\nu_k=k$ represents a stronger treatment effect over time and $\nu_k=0$ indicates no treatment effect. For each scenario of $\nu_k$ and each setting of $(s_A, \rho)$, we simulate 100 data replicates with sample size $n=500$.

Define the regime of always treat as $q_1$ and not treated as $q_0$. Relative to each simulated dataset, we aim to compare the overall mixed ATE \citep{li2022bayesian} at $t=2$ under the regimes of being always on treatment, $\overline{a}_{1:2}(q_1) = (1,1)$, versus not treated, $\overline{a}_{1:2}(q_0) = (0,0)$. By consistency and conditional sequential exchangeability assumptions, for any treatment regime $q$ of interest, i.e. $\overline{a}_{1:2}(q) = (a_1, a_2)$, target counterfactual quantity can be expressed as observed variables via g-formula as $\mathbb{E}[Y_{i2}(q)] = \mathbb{E}(Y_{i2}| A_{i1}=a_1, A_{i2}=a_2)$, derived in Appendix \ref{simg}. Population ATE, is the average difference between counterfactual outcomes $Y_{i2}(q_1)$ and $Y_{i2}(q_0)$, 
 \begin{align}\label{MATE}
     g(q_0, q_1) =&\mathbb{E}[Y_{i2}(q_1)] - \mathbb{E}[Y_{i2}(q_0)] \nonumber\\=& \mathbb{E}(Y_{i2}| A_{i1}=a_1, A_{i2}=a_2) - \mathbb{E}(Y_{i2}| A_{i1}=0, A_{i2}=0) \nonumber\\
     =&\sum^2_{k=1} \frac{k}{2}\times 1\{ a_1 + a_2=k\} + 0.4\times 0.5a_1.
 \end{align}
 %$g(\overline{a}_t) = \mathbb{E}(Y_{it} | \overline{A}_{i,1:t} = \overline{a}_{1:t})$. 
We estimate the population ATE by mixed ATE, $\widehat{g}(q_0, q_1) = \widehat{\mathbb{E}}[Y_{i2}(q_1)] - \widehat{\mathbb{E}}[Y_{i2}(q_0)]$, where we use sample-specific empirical distribution $\widehat{P}(V=v)$ instead of $P(V=v)$, the population distribution of $V$ involved in the derivation of population ATE.

By equation \eqref{MATE}, the true population ATE is $g(q_0, q_1) = 1.2$ for scenario $\nu_k = k$ and 0 for scenario $\nu_k = 0$. For each data replicate, we sample the posterior predictive distribution of the mixed ATE by marginalizing over $b^A_i \sim N(0,\widehat{s}^2_A)$ under $\widehat{s}_A \in \{0, 0.3, 1\}$, where $\widehat{s}_A$ represents the assumed degree of variation in the unexplained treatment assignment heterogeneity. Different combinations of simulation truth $(s_A, \rho)$ and assumed parameter $\widehat{s}_A$ explore the following three cases: (1) no unmeasured confounding ($\rho=0$), (2) unmeasured confounding exists with correctly specified models ($\rho \neq 0, \widehat{s}_A = s_A$), and (3) unmeasured confounding exists with a mis-specified extent of treatment assignment heterogeneity ($\rho \neq 0, \widehat{s}_A \neq s_A$). Note that the statement about unmeasured confounding is based on the assumptions encoded in the causal DAG and model choices. Let $N_{post}$ be the number of posterior draws. Given the $r$th data replicate under a simulation setting, we summarize the posterior samples of mixed ATE $(g^{(r)}_1,\ldots, g^{(r)}_{N_{post}})$ with its posterior mean $\overline{g}^{(r)} = \sum^{N_{post}}_{\ell=1} g^{(r)}_\ell / N_{post}$ and 95\% credible interval $(L^{(r)}, U^{(r)})$. We aggregate across simulation replicates by mean squared error (MSE) $\frac{1}{100} \sum^{100}_{r=1}[\overline{g}^{(r)} - g(q_0, q_1)]^2$ and coverage $\frac{1}{100} \sum^{100}_{r=1} \mathbbm{1}\{g(q_0, q_1) \in (L^{(r)}, U^{(r)})\}$. 

\begin{comment}
\begin{align*}
Y_{it} = &\beta^Y_0 + V_i\beta^Y_1 + t\beta^Y_2 + \sum^2_{k=1} 1\{ \sum^t_{s=1}A_{i,s}=k\}\beta^Y_{3k} + Y_{i,t-1}\beta^{Y}_4 + b^Y_{i0} + e^Y_{it}\\
=&0.4- 0.3V_i - 0.1 t +  \sum^2_{k=1} \frac{k}{2} \times 1\{ \sum^t_{s=1}A_{i,s}=k\}+0.4Y_{i,t-1}+ b^Y_{i0} +  e^Y_{it}\\
\text{logit}\{P(A_{it}(s) = 1 |&A_{i,t-1}(s)=0 )\} \\
	=& \beta^A_0 + V_i\beta^A_1 + t\beta^A_2 + Y_{i,t-1}\beta^A_4 +b^A_{i0}\\
=& -0.1V_i 	- 0.5 t -0.35 Y_{i,t-1} +b^A_{i0}
\end{align*}
Define $b_i = ( b^Y_{i0},  b^A_{i0})$, $g(\overline{a}_t) = \mathbb{E}(Y_{it} | \overline{A}_{it} = \overline{a}_t)$, $g(\overline{a}_t | V_i) = \mathbb{E}(Y_{it} | \overline{A}_{it} = \overline{a}_t, V_i)$, and $g(\overline{a}_t | V_i, b^A_i) = \mathbb{E}(Y_{it} | \overline{A}_{it} = \overline{a}_t, V_i, b^A_i)$. At each time point,
\begin{align*}
& g(1) - g(0) = \beta^Y_{31} = 0.5\\
& g(1| V_i) - g(0| V_i) = \beta^Y_{31} \\
& g(\{a_1,a_2\}) - g(\{0,0\}) \\
&\hspace{5em} = \sum^2_{k=1} \beta^Y_{3k} \times 1\{ a_1 + a_2=k\} + \beta^Y_{4}[ g(a_1) - g(0)] \\
&\hspace{5em} = \sum^2_{k=1} \frac{k}{2}\times 1\{ a_1 + a_2=k\} + 0.4\times 0.5a_1 \\
& g(\{a_1,a_2\}|  V_i) - g(\{0,0\}|  V_i)\\
&\hspace{5em} = \sum^2_{k=1} \beta^Y_{3k} \times 1\{ a_1 + a_2=k\} + \beta^Y_{4}[ g(a_1 | V_i) - g(0| V_i)] \\
&\hspace{9.5em} + \mathbb{E}(b^Y_{i0} | \overline{Y}_{i1}(a_1), V_i) -  \mathbb{E}(b^Y_{i0} | \overline{Y}_{i1}(0), V_i) \\
&\\
&  V_i \sim \text{Bernoulli}(0.5),  Y_{i0} \sim N(0,1)
\end{align*}
 \end{comment}

\begin{figure}[h!]
\centering
 \includegraphics[scale=0.75]{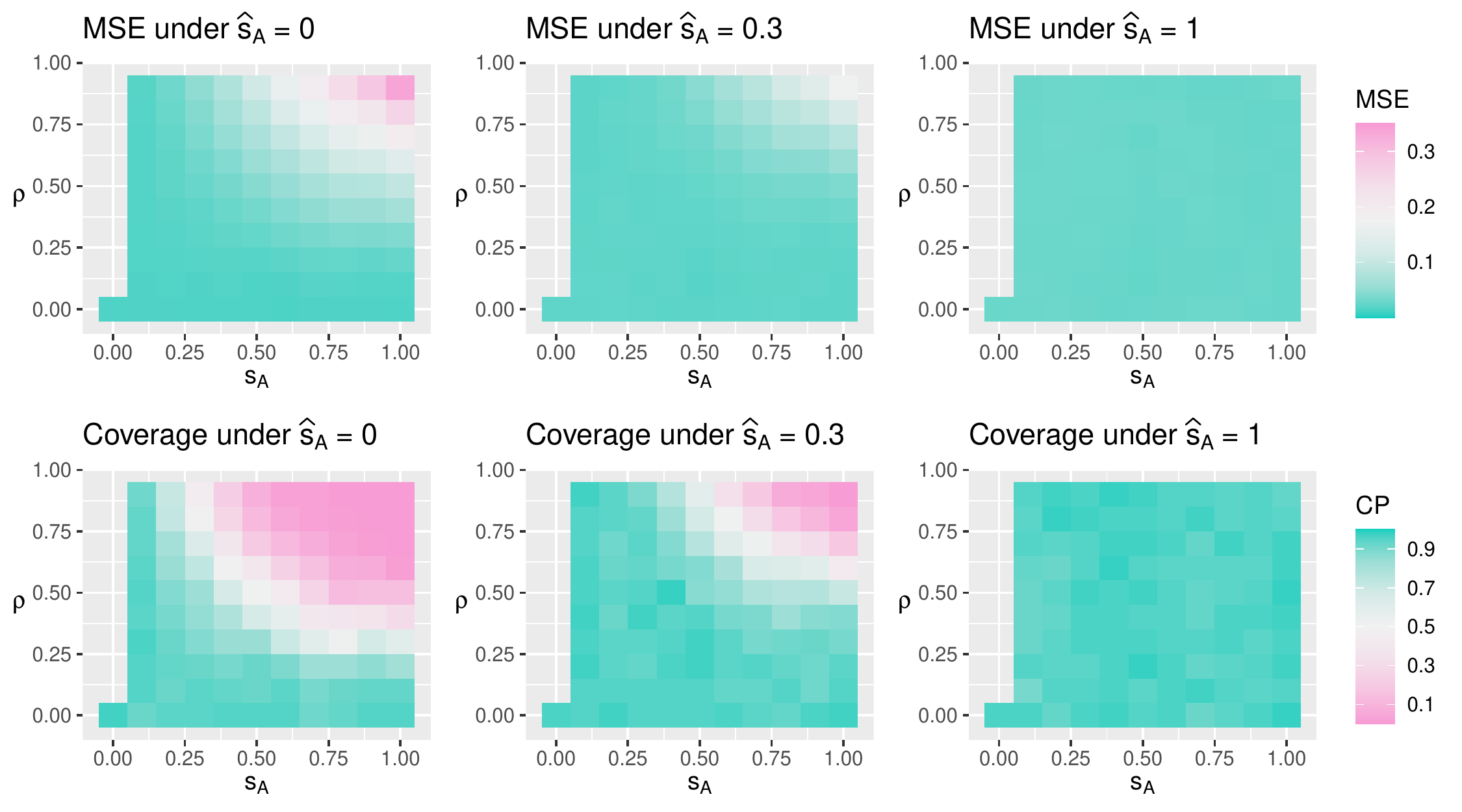}
        \caption{Under true treatment effect being 1.2 at the second time point, the figure displays mean squared error (MSE) and posterior coverage for mixed ATE under different simulation truth $(s_A, \rho)$ and assumed model parameter $\widehat{s}_A$. Color green refers to better estimation, e.g. lower MSE and higher coverage probability.}
\label{fig:sim1}
\end{figure}

\begin{figure}[h!]
\centering
\includegraphics[scale=0.95]{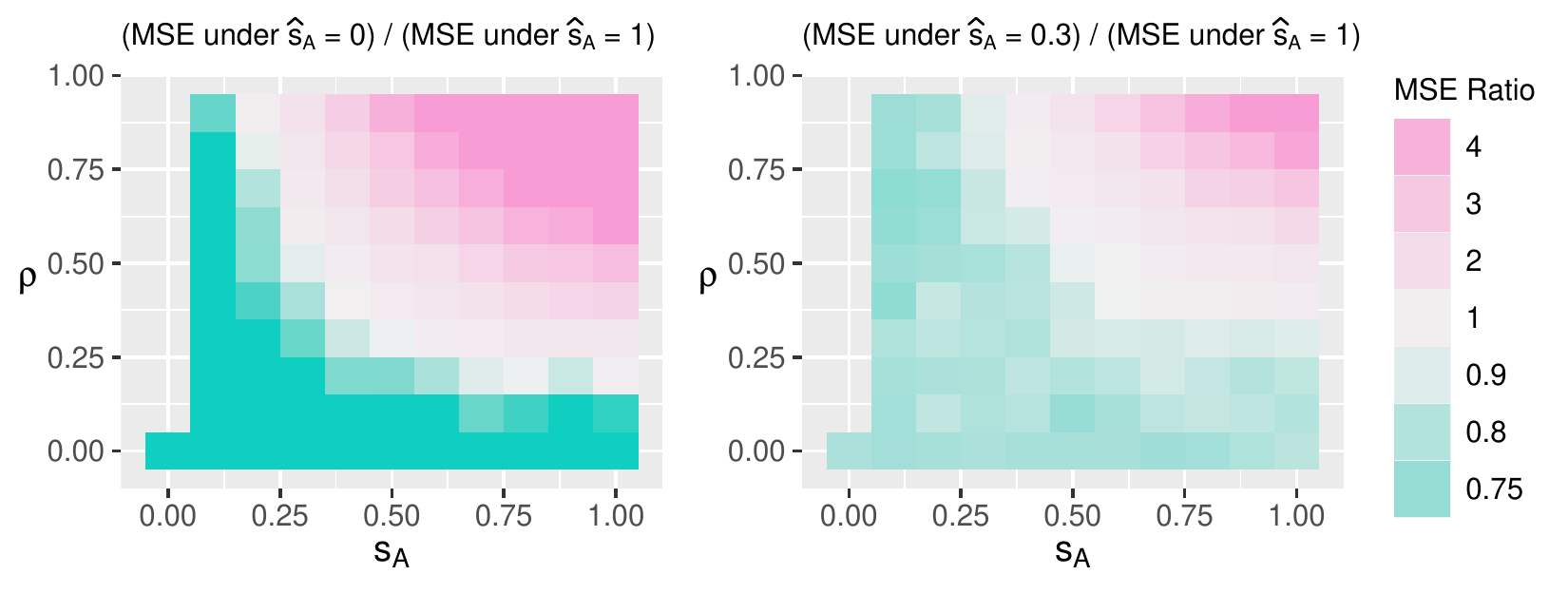}
        \caption{MSE ratio under true treatment effect being 1.2 at the second time point.}
\label{fig:MSEratio1}
\end{figure}

\begin{figure}[h!]
\centering
 \includegraphics[scale=0.75]{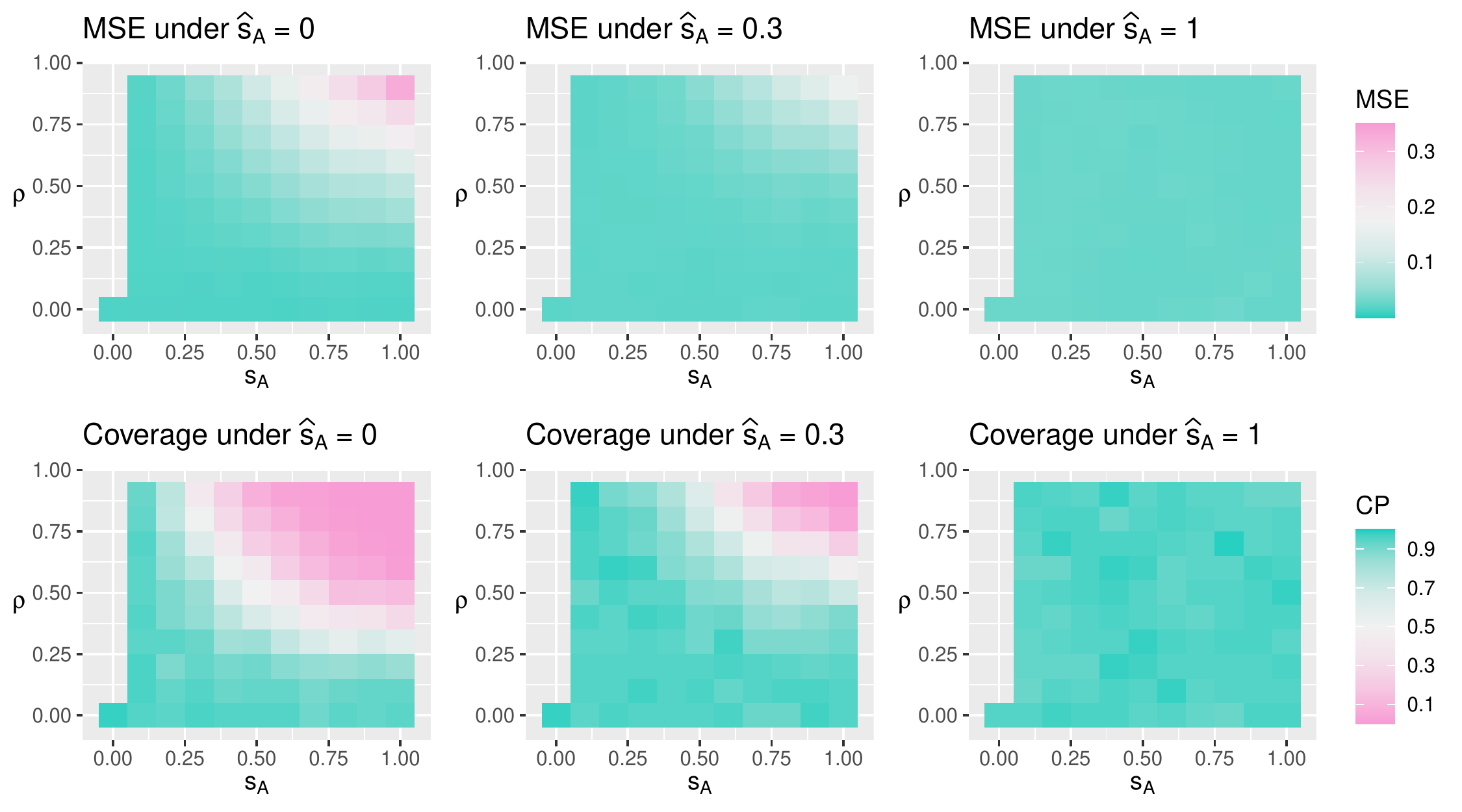}
        \caption{Under true treatment effect being 0 at the second time point, the figure displays mean squared error (MSE) and posterior coverage for mixed ATE under different simulation truth $(s_A, \rho)$ and assumed model parameter $\widehat{s}_A$. Color green refers to better estimation, e.g. lower MSE and higher coverage probability.}
\label{fig:sim0}
\end{figure}

\begin{figure}[h!]
\centering
\includegraphics[scale=0.95]{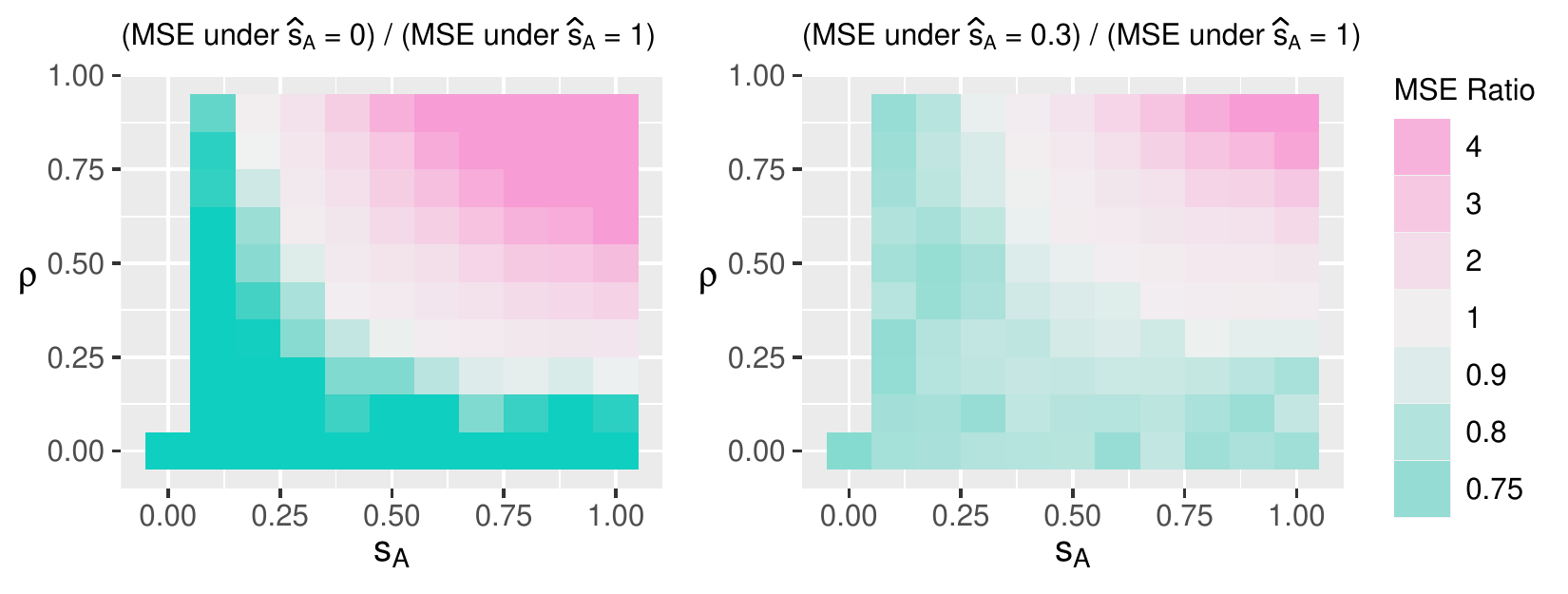}
        \caption{MSE ratio under no treatment effect.}
\label{fig:MSEratio0}
\end{figure}

For scenario $\nu_k = k$, 
Figure \ref{fig:sim1} displays the MSE and coverage of posterior mixed ATE in the first and second rows, respectively. The three columns from left to right correspond to estimations under $\widehat{s}_A$ being 0, 0.3, and 1, respectively. For each plot, the horizontal and vertical axes are the true parameters $s_A$ and $\rho$ under which data replicates were generated. Green indicates better estimation of the causal effect, i.e. lower MSE and higher posterior coverage. We observe that the causal effect is estimated relatively better when $\widehat{s}_A$ is no larger than the true value $s_A$. In addition, when there is no or close to no unmeasured confounding, i.e. $\rho$ is close to zero, the estimated mixed ATE is robust to the posited value $\widehat{s}_A$. In other words, poor estimation and coverage occur when the assumed variation in treatment assignment heterogeneity differs significantly from its true value and there is substantial unmeasured confounding. For $\widehat{s}_A=1$, estimation seems to be always better, so we also visualized in Figure \ref{fig:MSEratio1} the ratio of MSE under $\widehat{s}_A$ being 0 versus 1 and 0.3 versus 1. We see that estimations are better when the $\widehat{s}_A$ is no larger than the truth and when no unmeasured confounding is true, smaller values of $\widehat{s}_A$ are favored regardless of the true value $s_A$. For the other scenario, $\nu_k=0$, the treatment has no effect on the outcome at all times and $\mathbb{E}[Y_{it}(q)]$ does not depend on the treatment path. Figures \ref{fig:sim0} and \ref{fig:MSEratio0} summarize the results and yield similar conclusions as under $\nu_k = k$. When $\widehat{s}_A$ is close to the truth $(\widehat{s}_A\approx s_A)$ or when no unmeasured confounding is close to being true ($\rho \approx 0$), the method can find null-effect estimates well. 
%The g-null paradox \citep{mcgrath2021revisiting} implies that when there is no treatment effect and that past treatment affects time-varying confounders, estimating population ATE using parametric g-formula will in general lead to bias. The scenario of $\nu_k = 0$ meets the conditions of g-null paradox. 

 %Do it for two cases,  $g(\{a_1,a_2\}|  b^A_i) - g(\{0,0\}|  b^A_i)$ given the truth about $\{b^A_i; i = 1,\ldots, n\}$ Focus on $ g(\{1,1\}|  V_i) - g(\{0,0\}|  V_i)$. 

 %Assuming the existence of heterogeneity in the propensity of assigning treatment robustifies the estimation of MATE. That is, the MSE of treatment effect projections is similar when there is high degree of heterogeneity as assumed and when there is no or close to no heterogeneity, e.g. $\rho$ close to zero. In contrast, if we assume there is little or no heterogeneity, e.g. $\widehat{s}_A$ close to zero, we can get small MSE when the assumed $\widehat{s}_A$ is correct, but poor estimation and coverage of the MATE when the assumption of small heterogeneity is incorrect and when there is substantial unmeasured confounding.

\section{Application}

The proposed method is applied to clinical data from scleroderma patients collected longitudinally through the Johns Hopkins Scleroderma Center Research Registry. The application aims to study the causal effectiveness of MMF initiation regimes among the subgroup who were treated with MMF. The inference is performed by sampling and comparing the posterior predictive distribution of counterfactual outcome trajectories across time intervals under different treatment regimes, where outcomes are continuous time-varying multivariate biomarkers. Disease onset is defined by the emergence of symptoms, which typically occurs prior to and is inquired about during the enrollment visit.  Patients whose enrollment visits occurred within six years of disease onset and between 2010 and 2020 are included in the analysis data. The analysis utilizes individual clinical histories prior to February 28, 2022, and the observed maximum duration of follow-up in this data is ten years. Specifically, the data include all available follow-up visits when MMF was never taken, and up to two years after the first occurrence of continuous MMF use. The study includes 506 scleroderma patients who had not previously been treated with MMF at the time of enrollment, with 194 of them starting MMF during follow-up. Among these individuals, 80$\%$ are females, 20$\%$ are African Americans, and 40$\%$ have diffuse scleroderma; this is representative of the overall cohort. Age at disease onset has the first, second, and third quartiles as 38, 48, and 58 years old. Because patient visits are anticipated to be every six months, our analysis frames the progression of time-varying variables by six-month intervals. Over 90$\%$ of the observed MMF initiation happened during the first five post-enrollment time intervals.

To investigate the efficacy of MMF over the course of continuous use, assuming tolerance to the drug, we focus on the first two years of MMF usage in  patients who were treated with MMF. Suppose person $i$ was observed to start using MMF at time $\tilde{s}_i$, we evaluate the effectiveness of MMF  by comparing the regimes of continuously taking MMF versus no MMF use during the two-year period of time intervals $[\tilde{s}_i, \tilde{s}_i+4)$. In this study, we consider outcomes $Y_{it}$ to be measurements of the modified Rodnan skin score (mRSS) and lung scores evaluated by forced vital capacity percent predicted (FVC) and diffusing capacity for carbon monoxide percent predicted (DLCO). FVC and DLCO are continuous scores and are standardized for analysis. The skin score mRSS is a continuous measure with a range of 0 to 51 and we quantilized the mRSS to the standard normal distribution. Missingness in biomarker measurements is common due to the nature of clinical data being observed only when patients present for clinical visits. Among the included patients in this study, 82.6$\%$, 83.6$\%$, and 50$\%$ had at least one interval without measurement for FVC, DLCO, and mRSS, respectively. Visit patterns may potentially confound the effectiveness of MMF on the biomarkers. We define confounders $M_{it}$ to be indicators of whether FVC, DLCO, and mRSS were updated for person $i$ at time interval $t$, representing visit pattern over time.

\begin{figure}[!htp]
        \includegraphics[scale=1]{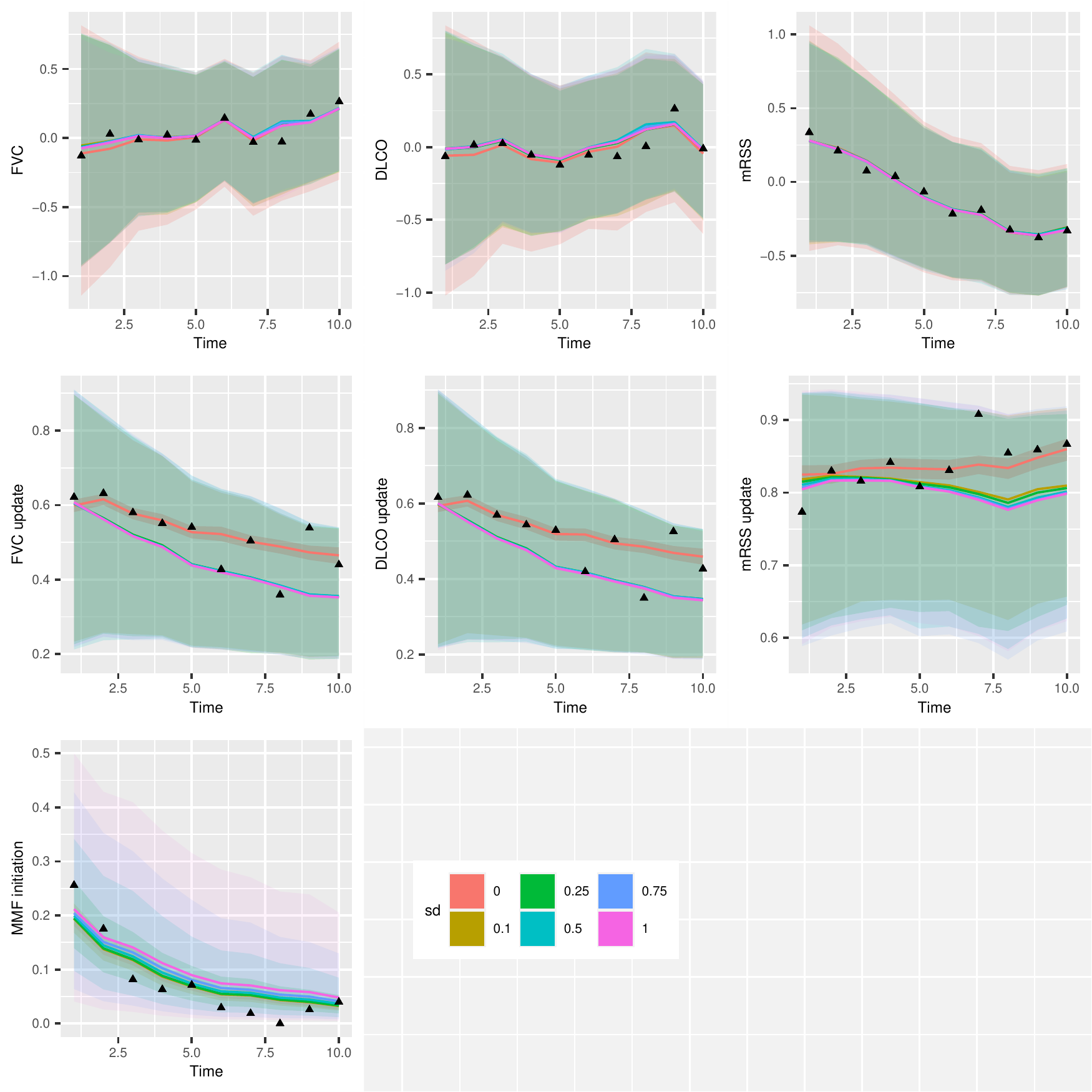}
        \caption{Application accuracy plot.}
\label{fig:acc}
\end{figure}

We define $V_i$ to be the vector of baseline demographic variables including sex, race, and age. Let $B_i$ be the baseline disease type, i.e. indicator of diffuse scleroderma. Based on domain knowledge, biomarkers are considered to progress upon time since disease onset, denoted as $S_{it} = t+O_i$, where $O_{i}$ is the duration between disease onset and the enrollment visit.  Write $\tilde{Y}_{it}$ as the carried forward measurement of biomarkers at time $t$.  Because we limit the study to no more than two years of continuous MMF use, i.e. four intervals, dosage information at time $t$ can be summarized by the vector $D(A_{i,0:t}) = \{\mathbbm{1}\big( \sum^t_{\ell=1}A_{i\ell}=1\big),\ldots,\mathbbm{1}\big( \sum^t_{\ell=1}A_{i\ell}=4\big)\}$, where $A_{i, 0:t} = (A_{i1},\ldots, A_{it})$ is the binary indicator vector of whether person $i$ was observed to be on MMF over time. In addition, time between disease onset and MMF initiation is denoted by $I_{it}$, which equals zero when $A_{it} = 0$ and equals $S_{it'}$ when $A_{it} = 1$, where $t'$ is the time of treatment initiation satisfying $A_{i,t'-1} = 0$, $A_{it'} = 1$, and $t' \le t$. The following joint model is assumed for outcomes, time-varying confounders, and treatment assignment,
\begin{align*}
Y_{it} | (M_{it} = 1)=& \phi_1(\mathcal{H}_{it}) \beta^Y_1 + \phi_2(\mathcal{H}_{it})\phi_A(\overline{A}_{i,0:t})^T\beta^Y_2 + b^Y_{i0} + e^Y_{it} \\
\text{logit}\{P(M_{it} = 1)\}=& \phi_1(\mathcal{H}_{it}) \beta^M_1 + \phi_2(\mathcal{H}_{it})\phi_A(\overline{A}_{i,0:t})^T\beta^M_2 + b^M_{i0} \\
\text{logit}\{P(A_{it} = 1 |A_{i,t-1}=0 )\} =& \phi_1(\mathcal{H}_{it}) \beta^A_1  + b^A_{i0}
\end{align*}
where 
$$(b^Y_{i0},b^M_{i0},b^A_{i0})^T\sim N(0,G),$$
$$\phi_1(\mathcal{H}_{it}) = \{1, \tilde{Y}_{i,t-1}, V_i, B_i, ns(S_{it},\nu_s), B_i \times ns(S_{it},\nu_s)\} ,$$
$$\phi_2(\mathcal{H}_{it})\phi_A(\overline{A}_{i,0:t})^T = \{ D(\overline{A}_{i,0:t}), V_i\times D(\overline{A}_{i,0:t}),B_i\times D(\overline{A}_{i,0:t}),  I_{it}\times D(\overline{A}_{i,0:t})\},$$
and we assume $\nu_s=4$. Note that both the outcomes $Y_{it}$ and confounders $M_{it}$ are multivariate, i.e. $Y_{it}$ and $M_{it} \in \mathbb{R}^3$. Specifically, $(b^Y_{i0},b^M_{i0},b^A_{i0})\in \mathbb{R}^7$ and the covariance matrix $G\in  \mathbb{R}^{7\times 7}$. Model validation results are summarized in Figure \ref{fig:acc}.
Black triangles represent the observed mean of time-varying variables at each time. The colored curves and areas represent the posterior mean and 95$\%$ posterior credible interval of the one-step forward prediction for each time-varying variable under various posited values of $b^A_{i0}$'s standard deviation. 
%Specifically, we can write the vector of random effects as  $b^Y_{i0} = (b^{Y_1}_{i0},b^{Y_2}_{i0},b^{Y_3}_{i0})$ and $b^M_{i0} = (b^{M_1}_{i0},b^{M_2}_{i0},b^{M_3}_{i0})$, 

\begin{figure}[htp]
        \includegraphics[scale=0.8]{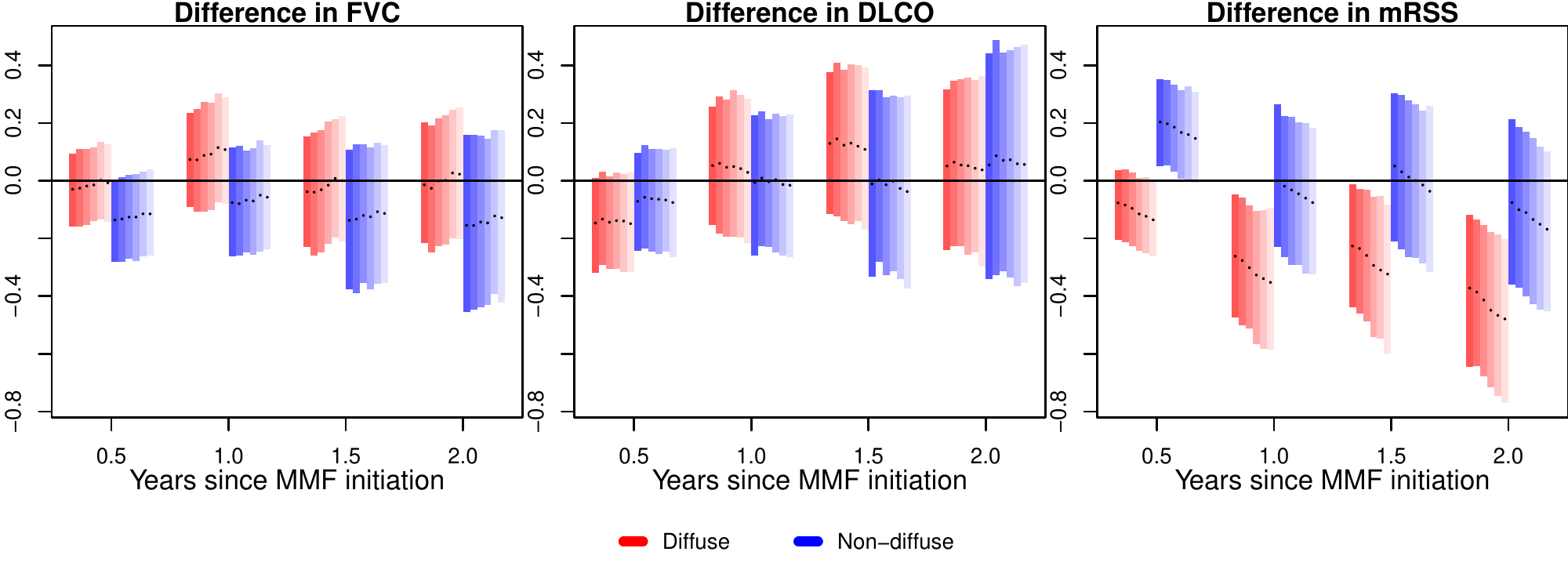}
        \caption{Application causal estimation by subgroup diffuse versus nondiffuse.}
\label{fig:cp_DN}
\end{figure}

For each person $i$ who was observed to have MMF during follow-up, we initiate the comparison of two regimes $q_1$ and $q_2$ at time $\tilde{s}_i$, conditional on the person's clinical history up to time $\tilde{s}_i-1$, i.e. $(V_i, \overline{A}_{i,0:(\tilde{s}_i-1)}, \overline{Y}_{i,0:(\tilde{s}_i-1)}, \overline{M}_{i,0:(\tilde{s}_i-1)})$.  Based on the algorithm outlined in Appendix \ref{appendix:pseudo}, we sample from the posterior predictive distribution of $(\overline{Y}_{i,\tilde{s}_i:(\tilde{s}_i+3)}(q_z),\overline{M}_{i,\tilde{s}_i:(\tilde{s}_i+3)}(q_z))$, which is the counterfactual trajectories under regime $q_z$, $z = 1,2$, and obtain posterior samples of the counterfactual trajectories, denoted as $\{ \overline{Y}^{(\ell)}_{i,\tilde{s}_i:(\tilde{s}_i+3)}(q_z), \overline{M}^{(\ell)}_{i,\tilde{s}_i:(\tilde{s}_i+3)}(q_z); \ell = 1,\ldots,N_{post})\}$. Causal comparative effectiveness between the two regimes is quantified by the averaged differences in biomarkers over time, $\mathbb{D}_j = \{ \sum_{i\in T} d^{(\ell)}_{ij} / N_{T}; \ell=1,\ldots,N_{post}\}$, where $j \in \{0,1,2,3\}$ indexes time since regimen application, $d^{(\ell)}_{ij} = Y^{(\ell)}_{i,\tilde{s}_i+j}(q_1) - Y^{(\ell)}_{i,\tilde{s}_i+j}(q_2)$, and $T$ is the subgroup of interest. Figure \ref{fig:cp_DN} displays the CMATE of MMF among the treated individuals and compares the two regimes, initiate MMF as observed versus no MMF. The figure depicts the posterior mean and 95$\%$ credible interval of $\mathbb{D}_j$ over the two years of regimen comparison, i.e. $j=0, 1, 2, 3$, under posited values of $\widehat{s}_A\in \{0, 0.1, 0.25, 0.5, 0.75, 1\}$ indicated by the decrease in opacity as $\widehat{s}_A$ increases, stratified by diffuse scleroderma status.

Figure \ref{fig:cp_DN} shows that incorporating MMF into the treatment of patients with diffuse scleroderma has a significant effect on skin score during the two years after drug initiation, assuming drug tolerance and continuous use of the drug. This is consistent with the Scleroderma Lung Study II, which found that MMF significantly improved mRSS in patients with diffuse scleroderma at the end of 24 months. We investigated the incorporation of MMF in the treatment of scleroderma patients, whereas the clinical trial compared the use of MMF alone for 24 months versus cyclophosphamide for 12 months followed by a placebo for 12 months. Furthermore, results indicate that there is no sufficient evidence that MMF improves FVC when averaged among the subgroups of individuals with diffuse or limited/sine scleroderma, given that patients may be receiving other therapies. Our findings also show that adding MMF to the treatment of patients who do not have diffuse scleroderma has no discernible benefit in the lung or skin within the two years of MMF initiation; this has clinical implications because MMF is an immunosuppressive agent that may increase the risk of serious infection and gastrointestinal side effects. Observe that drug combination and treatment practices for regimens containing or not containing MMF may differ in the real world; further research is needed to examine the impact of the difference in practices and treatment patterns resulting from the use or nonuse of MMF.

%Person fake ID 11 who initiated MMF at 1 year (2nd time interval) and had history of standardized FVC (0.7, -1.2) DLCO (-0.1, -1.2) MRSS (-0.5, -0.5). By assumption, $b^A_i\sim N(0, 0.1^2)$. Conditional on this person's history, Laplace approximation generates mean of $b^A_i$ having mean -0.0075 and IQR (-0.08, 0.06), sd of $b^A_i$ having mean 0.007 and IQR (0.006, 0.008), the simulated $b^A_i$ for this person then has mean -0.007 and IQR (-0.08, 0.06), range (-0.18,0.2). This is a narrower range than that of the same amount of samples randomly drawn from $N(0, 0.1^2)$, which is (-0.27, 0.28) from one run.

\section{Discussion}

Deciding which treatment regime is better for patients of a specific subgroup or history pattern is a basic question for treating patients in clinics. Causal inference is a natural tool for answering such questions, but characteristics of clinical data need to be accommodated for valid inference when evaluating the efficacy of treatment paths. Observational longitudinal clinical datasets often include treatment assignments that are not randomized based on observed patient history, as well as irregular measurements that may yield informative missingness from patients' visit patterns. A key factor in comparing treatment paths is the natural heterogeneity in treatment assignment and biomarker dynamics that goes beyond what observables can explain. These are typical features of longitudinal clinical datasets. Choosing the most effective treatment regimen for one type of patient requires summarizing evidence from a population of patients of a similar type while accounting for such person's specific biomarker trends. The statistical model used to answer this question is complex by nature. This paper describes the simplest possible model that accommodates these characteristics, such as nonrandom treatment assignment and patient heterogeneity, and provides a tool for the comparison of treatment paths.

The main contribution of this work is to develop a Bayesian framework for causal inference with observational longitudinal data, estimating the subgroup effectiveness of binary treatment paths on longitudinal outcomes while accounting for time-varying confounders and allowing the existence of time-invariant unmeasured confounding.
We propose to simultaneously model biomarker dynamics and treatment assignment by MGLMM, which retains the capability to deal with unmeasured confounding to some degree when the model specification is reasonably close to being correct. Since mixed-effects models do not rely on the assumption of no unmeasured confounding, which is required by the majority of the existing g-estimation methods, our approach gives the possibility of consistently estimating the causal effects of treatment paths even when unmeasured confounding certainly exists or when an unconfounded variable or instrumental variable is not available. We note that the MGLMM introduced here does not deal with time-varying unmeasured confounding. When a random slope for a time-varying variable is specified in the model, it is considered an unobserved trait that is time-invariant but characterizes patient heterogeneity in dynamic progression. Furthermore, MGLMM has a specific representation of unmeasured confounding, the degree of which is governed by the unexplained variation in treatment assignment $s_A$ and the correlation between the treatment assignment and biomarker dynamics that operates through the correlation parameter $\rho$. A small $\rho$ and a large $s_A$, or a large $\rho$ and a small $s_A$, may both lead to heavy unmeasured confounding. The method has the potential to be extended to guide the inclusion of other latent variable models in Bayesian causal inference.

Note that our proposal does rely on the parametric assumption about the joint distribution of outcomes, time-varying confounders, and treatment assignment. We cannot test the assumption of no unmeasured confounding unless the randomization of treatment assignment is guaranteed or controlled. Under the strong assumption of no unmeasured confounding, a strong parametric model assumption would not be required for valid causal inference. However, in the example of treating patients with chronic rare diseases in clinics, unmeasured confounders unavoidably exist. We recognize that there is no free lunch in causal inference and to relax the uncounfoundedness assumption, the tradeoff here is to make additional assumptions on model specifications, which is also largely untestable. A completely nonparametric causal effect in observational data cannot be identified and untestable assumptions are always needed for the identification of causal effects. The parametric model assumption facilitated the identification of treatment effects even when unmeasured confounders may exist. The structured unmeasured confounding represented in the MGLMM provides insight into how the very specific kind of unmeasured confounding impact the causal effect of treatment paths.

Method-wise, we adopt the Bayesian g-computation algorithm (GCA) by incorporating MGLMM as the time-evolving generative component, while accounting for the real-time update of subject-specific unobserved stable traits as patient history accumulates over time.  Our proposal makes subgroup evaluation of treatment paths possible by involving time-varying estimation of latent variables in the GCA, instead of marginalizing them out as in population ATE. Furthermore, the method provides a way of incorporating propensity scores (PS) in Bayesian causal inference. Existing ways of combining PS and outcomes models include specifying outcomes distribution based on PS, having shared parameters or priors between PS and outcome models, or using posterior-based inverse probability weighting or
doubly robust estimators \citep{li2022bayesian}. Our method falls under the category of having shared parameters or priors between PS and outcome models, using a multivariate Gaussian latent structure to connect them through covariance between latent variables. Lastly, our method provides an alternative way to assess sensitivity analysis in causal inference. Instead of assuming no unmeasured confounding and conducting post hoc analysis to assess bias due to unobserved confounding, the proposal estimates causal effects conditional on different posited values of the sensitivity parameter, which is the variance of unobserved treatment assignment heterogeneity. Future extensions of the method should
consider categorical and count outcomes, multiple or continuous treatments, and more flexible distributional assumptions on the patient heterogeneities. 

\section*{Acknowledgment}

This work was supported in part by the Johns Hopkins inHealth initiative, the Scleroderma Research Foundation, the Nancy and Joachim Bechtle Precision Medicine Fund for Scleroderma, the Manugian Family Scholar, the Donald B. and Dorothy L. Stabler Foundation, the Chresanthe Staurulakis Memorial Fund, the Sara and Alex Othon Research Fund, NIH P30AR070254 and NIH/NIAMS K24AR080217.

We thank Adrianne Woods for excellent database management support, the patients of the Johns Hopkins Scleroderma Center Research Registry for their participation in this foundational resource, and Dr. Joseph Hogan for his knowledgeable comments.

\newpage
\bigskip
\noindent\makebox[\linewidth]{\rule{\paperwidth}{0.4pt}}

\bigskip

\baselineskip 22pt

\bibliographystyle{apacite}

 \bibliography{ref22-11-12}

\section*{Figures}

\newpage

\begin{comment}
\begin{figure}[htp]
        \includegraphics[scale=0.8]{Causal_subgroup_band_conference_c_BW.pdf}
        \caption{Application causal estimation by subgroup white versus nonwhite.}
\label{fig:cp_BW}
\end{figure} 
\end{comment}

\newpage

\begin{comment}
\section{Tables}
\begin{table}[h!]
\centering
\begin{tabular}{ccccccc}
\hline
                 &   & n   & percent & 25\% & 50\% & 75\% \\
Male             & 0 & 411 & 0.8     &      &      &      \\
                 & 1 & 95  & 0.2     &      &      &      \\
African American & 0 & 412 & 0.8     &      &      &      \\
                 & 1 & 94  & 0.2     &      &      &      \\
Onset Age        &   & 506 & 100     & 38   & 48   & 58   \\
Diffuse          & 0 & 289 & 0.6     &      &      &      \\
                 & 1 & 217 & 0.4     &      &      &   \\ \hline  
\end{tabular}
\caption{Summary table of baseline variables.}
\label{tab:V}
\end{table}
\end{comment}

\newpage
\begin{appendices}

\section{Identification of the G-formula} \label{appendix:Gformula}

For simplicity, we ignore the subscript $i$ for indexing subjects. Assuming $\overline{A}_{0:h} = \overline{a}_{0:h}(q)$ and time-invariant latent treatment heterogeneity $b^A_i = b^A_i$, the distribution of counterfactual  trajectories for the future $\tau$ time intervals conditional on observed information up to time $h$ can be processed as follows.
\begin{align*}
&P(\overline{Y}_{(h+1):(h+\tau)}(q),\overline{M}_{(h+1):(h+\tau)}(q) | V, \overline{A}_{0:h}, \overline{Y}_{0:h}, \overline{M}_{0:h}, b^A_i )\\
& \text{by positivity and exchangeability,}\\
=& P(\overline{Y}_{(h+1):(h+\tau)}(q),\overline{M}_{(h+1):(h+\tau)}(q)  | V, \overline{A}_{0:h}, A_{h+1} = a_{h+1}(q), \overline{Y}_{0:h}, \overline{M}_{0:h}, b^A_i ) \\
& \text{by consistency,}\\
=& P(Y_{h+1}, M_{h+1} | V, \overline{A}_{0:(h+1)} = \overline{a}_{0:(h+1)}(q), \overline{Y}_{0:h},\overline{M}_{0:h}, b^A_i ) \\
&P(\overline{Y}_{(h+2):(h+\tau)}(q),\overline{M}_{(h+2):(h+\tau)}(q)| V, \overline{A}_{0:(h+2)} = \overline{a}_{0:(h+2)}(q), \overline{Y}_{0:(h+1)},\overline{M}_{0:(h+1)}, b^A_i )  \\
&\text{by induction,}\\
=& \prod^{h+\tau-1}_{s = h} P( Y_{s+1} , M_{s+1}| V, \overline{A}_{0:(s+1)} = \overline{a}_{0:(s+1)}(q), \overline{Y}_{0:s}, \overline{M}_{0:s}, b^A_i)  \\
& \text{account for and marginalize over patient heterogeneity,}\\
=&\prod^{h+\tau-1}_{s = h}  \int_{u_s} \int_{v_s} P( Y_{s+1},M_{s+1} | V, \overline{A}_{0:(s+1)} = \overline{a}_{0:(s+1)}(q), \overline{Y}_{0:s}, \overline{M}_{0:s}, b^Y = u_s, b^M = v_s,b^A_i)  \\
& \hspace{6em}  P(b^Y = u_s, b^M = v_s| V, \overline{A}_{0:(s+1)} = \overline{a}_{0:(s+1)}(q), \overline{Y}_{0:s}, \overline{M}_{0:s}, b^A_i) du_s dv_s\\
& \text{because counterfactual treatment path does not inform heterogeneity estimation,}\\
=&\prod^{h+\tau-1}_{s = h}  \int_{u_s} \int_{v_s} P( Y_{s+1},M_{s+1} | V, \overline{A}_{0:(s+1)} = \overline{a}_{0:(s+1)}(q), \overline{Y}_{0:s}, \overline{M}_{0:s}, b^Y = u_s, b^M = v_s,b^A_i)  \\
& \hspace{6em}  P(b^Y = u_s, b^M = v_s| V, \overline{A}_{0:h}, \overline{Y}_{0:s}, \overline{M}_{0:s}, b^A_i) du_s dv_s\\
& \text{by distributional assumptions illustrated in Figure \ref{fig:diagram},}\\
=&\prod^{h+\tau-1}_{s = h}   \int_{u_s} \int_{v_s}  P(Y_{s+1} | V, \overline{A}_{0:(s+1)} = \overline{a}_{0:(s+1)}(q), \overline{Y}_{0:s}, \overline{M}_{0:s}, b^Y = u_s)\\
& \hspace{6em}  P(M_{s+1} | V, \overline{A}_{0:(s+1)} = \overline{a}_{0:(s+1)}(q), \overline{Y}_{0:s}, \overline{M}_{0:s}, b^M = v_s)\\
& \hspace{6em}  P(b^Y = u_s, b^M = v_s| V,\overline{A}_{0:h}, \overline{Y}_{0:s}, \overline{M}_{0:s}, b^A_i)  du_s dv_s\\
&\text{parameterizing  MGLMM as linear models, we get}\\
=&\prod^{h+\tau-1}_{s = h}   \int_{u_s} \int_{v_s}   P(Y_{s+1} | V, \overline{A}_{0:(s+1)} = \overline{a}_{0:(s+1)}(q), \overline{Y}_{0:s}, \overline{M}_{0:s}, b^Y=u_s; \beta^Y, \sigma^2) \\
& \hspace{6em}P(M_{s+1} | V, \overline{A}_{0:(s+1)} = \overline{a}_{0:(s+1)}(q), \overline{Y}_{0:s}, \overline{M}_{0:s}, b^M=v_s; \beta^M)\\
& \hspace{6em}  P(b^Y = u_s, b^M = v_s| V,\overline{A}_{0:h}, \overline{Y}_{0:s}, \overline{M}_{0:s}, b^A_i; G)  du_s dv_s
\end{align*}

Given $b^A_i$, the g-formula for a conditional subgroup ATE is defined as a conditional mean of the potential outcome at the end of follow-up at time $h+\tau$ under a user-specified regime $q$. It can then be derived as below, %where baseline variables are assumed to be not informative on the distribution on the random effects, e.g. $(b^Y, b^M, b^A_i)| V, Y_0, M_0 \sim N(0, G)$.
\begin{align*}
   & \mathbbm{E}(Y_{h+\tau}(q)|V, \overline{A}_{0:h}, \overline{Y}_{0:h}, \overline{M}_{0:h}, b^A_i)\\
   =& \int_{y_\tau}y_\tau P(Y_{h+\tau}(q) = y_\tau|V, \overline{A}_{0:h}, \overline{Y}_{0:h}, \overline{M}_{0:h}, b^A_i)dy_\tau \\
   = & \int_{y_{h+\tau}}\int_{m_{h+\tau}}\ldots  \int_{y_{h+1}}\int_{m_{h+1}} \\
   & \hspace{1em}y_\tau P(\overline{Y}_{(h+1):(h+\tau)}(q) = \overline{y}_{(h+1):(h+\tau)}, \overline{M}_{(h+1):(h+\tau)}(q) = \overline{m}_{(h+1):(h+\tau)} | V, \overline{A}_{0:h}, \overline{Y}_{0:h}, \overline{M}_{0:h}, b^A_i)\\
   &\hspace{25em}  dm_{h+1} dy_{h+1}\ldots d m_{h+\tau} d y_{h+\tau}\\
   = & \int_{y_{h+\tau}}\int_{m_{h+\tau}}\ldots  \int_{y_{h+1}}\int_{m_{h+1}} y_\tau \bigg\{\prod^{\tau-1}_{s = 0}   \int_{u_s} \int_{v_s}  \\
   & \hspace{4em}P(Y_{s+1} =y_{s+1}| V, \overline{A}_{0:(s+1)} = \overline{a}_{0:(s+1)}(q), \overline{Y}_{0:s}, \overline{M}_{0:s}, b^Y=u_s; \beta^Y, \sigma^2) \\
& \hspace{4em}P(M_{s+1}=m_{s+1} | V, \overline{A}_{0:(s+1)} = \overline{a}_{0:(s+1)}(q), \overline{Y}_{0:s}, \overline{M}_{0:s}, b^M=v_s; \beta^M)\\
& \hspace{4em}  P(b^Y = u_s, b^M = v_s| V,\overline{A}_{0:h}, \overline{Y}_{0:s}, \overline{M}_{0:s}, b^A_i; G)  du_s dv_s\bigg\}  \\  &\hspace{25em}  dm_{h+1} dy_{h+1}\ldots d m_{h+\tau} d y_{h+\tau}\\
 = & \int_{y_{h+\tau}}\int_{m_{h+\tau}}  \int_{u_{\tau-1}} \int_{v_{\tau-1}}\ldots   \int_{y_{h+1}}\int_{m_{h+1}}  \int_{u_0} \int_{v_0}   \\
   & \hspace{4em}y_\tau \bigg\{\prod^{\tau-1}_{s = 0} P(Y_{s+1} =y_{s+1}| V, \overline{A}_{s+1} = \overline{a}_{s+1}(q),  \overline{Y}_s =\overline{y}_s, \overline{M}_s=\overline{m}_s, b^Y=u_s; \beta^Y, \sigma^2) \\
& \hspace{4em}P(M_{s+1}=m_{s+1} | V, \overline{A}_{s+1} = \overline{a}_{s+1}(q), \overline{Y}_s=\overline{y}_s, \overline{M}_s=\overline{m}_s, b^M=v_s; \beta^M)\\
& \hspace{4em}  P(b^Y = u_s, b^M = v_s| V,\overline{A}_{0:h}, \overline{Y}_{0:s}, \overline{M}_{0:s}, b^A_i; G)  \bigg\}\\
&\hspace{20em} du_0 dv_0dm_{h+1} d y_{h+1}\ldots du_{\tau-1} dv_{\tau-1} d m_{h+\tau} d y_{h+\tau}
\end{align*}

%Given that the target population are subjects who were not treated at baseline ($A_0=0$), we leave out $A_0=0$ from the following derivation. 
The population ATE conditional on $b^A_i$ can be obtained by further integrating over the distribution of observed clinical history in the target population,
\begin{align*}
   & \mathbbm{E}(Y_{h+\tau}(q)|b^A_i)=\int_v\int_{y_h}\int_{m_h} \cdots \int_{y_0}\int_{m_0} \mathbbm{E}(Y_{h+\tau}(q)|V=v, \overline{A}_{0:h}=\overline{a}_{0:h}, \overline{Y}_{0:h}=\overline{y}_{0:h}, \overline{M}_{0:h}=\overline{m}_{0:h}, b^A_i)\\
   & \hspace{6em}P(V=v, \overline{A}_{0:h}=\overline{a}_{0:h}, \overline{Y}_{0:h}=\overline{y}_{0:h}, \overline{M}_{0:h}=\overline{m}_{0:h}) dm_0dy_0\ldots dm_hdy_h dv
\end{align*}

The CMATE is computed as follows by substituting the target population distribution of the observable with the corresponding empirical distribution, $\widehat{P}(V=v, \overline{A}_{0:h}=\overline{a}_{0:h}, \overline{Y}_{0:h}=\overline{y}_{0:h}, \overline{M}_{0:h}=\overline{m}_{0:h})$.
\begin{align*}
   & \widehat{\mathbbm{E}}(Y_{h+\tau}(q)|b^A_i)=\int_v\int_{y_h}\int_{m_h} \cdots \int_{y_0}\int_{m_0} \mathbbm{E}(Y_{h+\tau}(q)|V=v, \overline{A}_{0:h}=\overline{a}_{0:h}, \overline{Y}_{0:h}=\overline{y}_{0:h}, \overline{M}_{0:h}=\overline{m}_{0:h}, b^A_i)\\
   & \hspace{6em}\widehat{P}(V=v, \overline{A}_{0:h}=\overline{a}_{0:h}, \overline{Y}_{0:h}=\overline{y}_{0:h}, \overline{M}_{0:h}=\overline{m}_{0:h}) dm_0dy_0\ldots dm_hdy_h dv.
\end{align*}

Heterogeneity in treatment assignment, $b^A_i$, is assumed to be marginally $N(0,v)$ in the target population. The  marginal mixed population ATE can then be obtained by integrating $b^A_i$ over its distribution $P(b^A_i=w)$ as $\widehat{\mathbbm{E}}(Y_{h+\tau}(q)) = \int_w \widehat{\mathbbm{E}}(Y_{h+\tau}(q)|b^A_i=w)P(b^A_i=w) dw$.

\newpage

\newpage

\section{Sequential Update of Random Effects} \label{appendix:sequential}

Without loss of generality, assuming Gaussian distribution and logit model for continuous and binary variables, respectively, the structural model can be written as follows ,
\begin{align*}
&Y_{it}|M_{it}=1 \sim  \eta^{Y}_{it}+ \sigma \psi^Y,\\
&P(M_{it}=1) = \frac{\exp(\eta^{M}_{it})}{1+\exp(\eta^{M}_{it})},\\
&P(A_{it}=1|A_{i,t-1}=0) = \frac{\exp(\eta^A_{it})}{1+\exp(\eta^A_{it})},
\end{align*}
where $\psi^Y \sim N(0,1)$, $\eta^{Y}_{it} = \eta^{Y}(\mathcal{F}_{it}, b^{Y}_{i};\theta^{Y})$, $\eta^{M}_{it} = \eta^{M}(\mathcal{F}_{it}, b^{M}_{i};\theta^{M})$, and $\eta^A_{it} = \eta^{A}(\mathcal{F}^A_{it}, b^{A}_{i};\theta^{A})$. %$\eta^{Y}_{it} = X_{it}\beta^{Y}+b^{Y}_{i0}$, $\eta^{M}_{it} = X_{it}\beta^{M}+b^{M}_{i0}$, and $\eta^A_{it} = X_{it}\beta^A+b^A_i_{i0}$.

Sequential update for random effects is implemented for each individual, conditional on biomarker dynamics up to time $t$ and observed treatment sequence up to time $h$, where $h \le t$. For the observed trajectories of subject $i$, the joint likelihood is
\begin{align*}
&P(\overline{Y}_{i,0:t},\overline{M}_{i,0:t},\overline{A}_{i,0:h}|b_i, \beta, \sigma)\\
&\propto \prod^{t}_{j=1} \bigg[\bigg(\frac{1}{\sigma} \exp\{-\frac{1}{2\sigma^2} (Y_{ij} - \eta^{Y}_{ij})^2\} \bigg)^{M_{ij}} \frac{\exp\{\eta^{M}_{ij}M_{ij}\}}{1+\exp(\eta^{M}_{ij})} \bigg]
\times  \prod^{h}_{j'=1}  \bigg[\frac{\exp\{\eta^A_{ij'}A_{ij'}\}}{1+\exp(\eta^A_{ij'})}\bigg]^{\mathbbm{1}(j'\le s_i)} ,
\end{align*}

where $s_i$ is the observed treatment initiation time for subject $i$, and the random effect $b_i$ has prior 
\begin{align*}
&P(b_i|G) \propto |G|^{-1/2}\exp(-\frac{1}{2} b^T_iG^{-1}b_i).
\end{align*}

The log posterior of $b_i$ can then be written as
\begin{align*}
&\log P(b_i| \overline{Y}_{i,0:t},\overline{M}_{i,0:t},\overline{A}_{i,0:h},  \beta, \sigma, G)\\
&\propto -\frac{1}{2}b^T_i G^{-1}b_i + \sum^{t}_{j=1}\bigg\{ -\frac{M_{ij}}{2\sigma^2} (Y_{ij} - \eta^{Y}_{ij})^2  + \eta^{M}_{ij}M_{ij} - \log [1+\exp(\eta^{M}_{ij})] \bigg\}\\
&\qquad + \sum^{min(h,s_i)}_{j'=1} \bigg\{\eta^{A}_{ij'}A_{ij'} - \log[1+\exp(\eta^{A}_{ij'})] \bigg\}.  %\mathbbm{1}(j\le s_i)
\end{align*}

 Using algorithms for constructing sampling chains, such as MCMC, in sampling $b_i$ would consume a significant amount of computational resources due to the complexity of calculating counterfactual individual trajectories. We consider a Laplace approximation of the  posterior distribution of $b_i$ for an easier posterior sampling. The mean of the approximated distribution is obtained by solving the following equation for a posterior mode $\hat{b}_i = (\hat{b}^Y_i,\hat{b}^M_i,\hat{b}^A_i)$,
$$\frac{\partial}{\partial b_i} \log P(b_i| \overline{Y}_{i,0:t},\overline{M}_{i,0:t},\overline{A}_{i,0:h},  \beta, \sigma, G)\bigg|_{b_i = \hat{b}_i} = 0,$$
where 
$$\frac{\partial}{\partial b_i} \log P(b_i| \overline{Y}_{i,0:t},\overline{M}_{i,0:t},\overline{A}_{i,0:h},  \beta, \sigma, G) 
= - G^{-1} b_i + 
\begin{pmatrix}
\sum^{t}_{j=1}\frac{M_{ij}}{\sigma^2_1} (Y_{ij} - \eta^{Y}_{ij})\\
\sum^{t}_{j=1} M_{ij} - \frac{\exp(\eta^{M}_{ij})}{1+\exp(\eta^{M}_{ij})}\\ 
\sum^{min(h,s_i)}_{j=1} A_{ij} - \frac{\exp(\eta^{A}_{ij})}{1+\exp(\eta^{A}_{ij})}
\end{pmatrix}.
$$
The variance of the approximated distribution is the asymptotic variance of $\hat{b}_i$, which is the inverse of the observed Fisher information matrix defined as follows
$$V = \bigg[- \frac{\partial^2}{\partial b_i \partial b^T_i}  \log P(b_i| \overline{Y}_{i,0:t},\overline{M}_{i,0:t},\overline{A}_{i,0:h},  \beta, \sigma, G)  \bigg|_{b_i = \hat{b}_i} \bigg]^{-1},$$
where 
\begin{align*}
& \frac{\partial^2}{\partial b_i \partial b^T_i}  \log P(b_i| \overline{Y}_{i,0:t},\overline{M}_{i,0:t},\overline{A}_{i,0:h},  \beta, \sigma, G) \\
=& -G^{-1} -  \text{diag}\bigg\{\frac{1}{\sigma^2} \sum^{t}_{j=1}M_{ij},
\sum^{t}_{j=1} \frac{\exp(\eta^{M}_{ij})}{[1+\exp(\eta^{M}_{ij})]^2},
\sum^{min(h,s_i)}_{j=1} \frac{\exp(\eta^{A}_{ij})}{[1+\exp(\eta^{A}_{ij})]^2}\bigg\}.
\end{align*}
As a result, an approximation to the posterior distribution $P(b_i| Y_{i,0:t},M_{i,0:t},A_{i,0:h},  \beta, \sigma, G)$ is the multivariate Gaussian distribution $MNV(\hat{b}_i,V)$.

%In our framework, we condition on the treatment heterogeneity $b^A_i$ being a specific value for each Bayesian projection of counterfactual biomarker dynamics in order to block information flow from fixing the treatment sequence to $\overline{a}_{0:t}(q)$ under a regime $q$ of interest. Therefore we fix the value of $b^A_i$ which creates identifiability, and estimates the causal effect for ranges of value of $b^A_i$. This is like fixing $b^A_i$ and lookin at the sensitivity of the causal effect to the unidentifiability of treatment preference.

Sequential update of counterfactual trajectories is also conditinoal on $b^A_i$ being a constant, i.e. $b^A_i=c$. We sequentially update the heterogeneity in biomarker dynamics conditional on history $(Y_{i,0:t},M_{i,0:t},A_{i,0:h})$, population level estimates $(\beta, \sigma, G)$, and $b^{A}_i = c$ as follows
$$(b^{Y}_i, b^{M}_i | b^{A}_i = c) \sim MNV(b_{\cdot|A}, V_{\cdot|A})$$
such that 
\begin{align}\label{bcond}
    b_{\cdot|A} = \begin{pmatrix}
\hat{b}^{Y}\\ 
\hat{b}^{M}
\end{pmatrix} + 
\begin{pmatrix}
V^{Y,A}\\ 
V^{M,A}
\end{pmatrix} (V^A)^{-1} (c-\hat{b}^A)
\end{align}
\begin{align}\label{Vcond}
   V_{\cdot|A} = \begin{pmatrix}
V^{Y} & V^{Y,M} \\ 
 & V^{M} 
\end{pmatrix} - \begin{pmatrix}
V^{Y,A}\\ 
V^{M,A}
\end{pmatrix} (V^A)^{-1} (V^{Y,A},V^{M,A}). 
\end{align}

%\begin{comment}
 Suppose we are simulating the counterfactual progression of patient's longitudinal measures with treatment sequence fixed as $\overline{a}^q_{0:t}$ under regime $q$, where the sequence up to time $h$ is the observed treatment, i.e. $A_{i,0:h} = \overline{a}^q_{0:h}$. If we write the third row of $G^{-1}$ as $(C_1,C_2,C_3)$, then the derivative entry relative to $b^A_i$ leads to
 \begin{equation}\label{bAdir}
     \sum^{min(h,s_i)}_{j=1} a_j - \frac{\exp(X_{ij}\beta^A+b^A_{i0})}{1+\exp(X_{ij}\beta^A+b^A_{i0})}  = C_1b^A_i + C_2 b^{M}_i+ C_3 b^{Y}_i,
 \end{equation}
and we can see that the specification of the counterfactual treatment sequence $\overline{a}_{(h+1):t}$ does not affect the estimation of $\hat{b}_i$. Note that $\sum^{min(h,s_i)}_{j=1} a_j $ is either 0 or 1, because the summation stops at the time of initiation. In the application, we focus on studying the effect of treatment initiation among those who were not treated before a time $h$, i.e. $h < s_i$ and $\sum^{min(h,s_i)}_{j=1} a_j = 0$. Hence, for the estimation of $\hat{b}_i$, equation \eqref{bAdir} imposes condition $ - \frac{\exp(X_{ij}\beta^A+b^A_{i0})}{1+\exp(X_{ij}\beta^A+b^A_{i0})}  = C_1b^A_i + C_2 b^{M}_i+ C_3 b^{Y}_i$, using only treatment information before an treatment initiation.
%\end{comment}

\section{Pseudocode for Generating Counterfactual Trajectories}\label{appendix:pseudo}

\begin{algorithm}[H]
\renewcommand{\thealgorithm}{}
	\small
	\caption{\textbf{for Dynamic Projection of Counterfactual Trajectories  under MGLMM}}%Update of Individual Trajectories
	\label{A1}
	%\begin{algorithmic}[1]
	\begin{algorithmic}
	\State  Conditional on:
	\State \hspace{1em} (a) observed history up to time $h$, $ (V_i,\overline{Y}_{i,0:h},\overline{M}_{i,0:h},\overline{A}_{i,0:h})$ 
	\State \hspace{1em} (b) posteriors of $(\theta^Y,\theta^M, \theta^A,G)$
	\State \hspace{1em} (c) $var(b^A_i) = v$, 
	\State Goal: make posterior predictive inference of $(\overline{Y}_{(h+1):T}(q),\overline{M}_{(h+1):T}(q))$ under regime $q$.
	%$\overline{a}_{0:T}$ where $a_0=0$ and $\overline{a}_{0:h}=\overline{A}_{0:h}$,
	%Given $(\overline{Y}_{1:h},\overline{M}_{1:h})$, baseline information $(V,Y_0)$, posterior estimate of population parameters $(\beta^Y,\beta^M, \beta^A, G, R)$, and treatment regime of interest $\overline{a}_{0:T}$ where $a_0=0$ and $\overline{a}_{0:h}=\overline{A}_{0:h}$,
		\State \textbf{Step 0}: Initialization 
		\State \hspace{1em}(a) draw subject-specific stochastic matrices  $\psi^Y, \psi^M\in \mathbbm{R}^{N_{post}\times (T-h)}$, $\psi^Y\sim \mathcal{N}(0,1)$ and $\psi^M\sim \mathcal{U}(0,1)$
		\State \hspace{1em}(b) $\mathcal{F}^{(\ell)}_{i,h+1}(q) = (V_i,\overline{Y}_{i,0:h},\overline{M}_{i,0:h},\overline{A}_{i,0:h},a_{h+1}(q)) $   for all $\ell$
      \State \hspace{1em}(c) for each $\ell$, draw $b^{A(\ell)}_i \sim f(b^A| V_i,\overline{Y}_{i,0:h},\overline{M}_{i,0:h},\overline{A}_{i,0:h}; \theta^{Y(\ell)},\theta^{M(\ell)},\theta^{A(\ell)},G^{(\ell)})$ if $h > 0$, \\otherwise draw $b^{A(\ell)}_i \sim N(0,v)$ 
		\State \hspace{1em}(d) $l=0$
	\While{$\ell < N_{post}$}
		\State \hspace{1em} \textbf{for} $t\in h+1,\ldots,T$ \textbf{do}
		
		\State \hspace{2.3em}\textbf{Step 1}: Calculate $(\hat{b}^{(\ell)}_i(q), V^{(\ell)}_i(q))$ conditional on $(V_i,\overline{Y}^{(\ell)}_{i,0:(t-1)}(q),\overline{M}^{(\ell)}_{i,0:(t-1)}(q),\overline{A}_{i,0:h})$
		
		\State \hspace{2.3em}\textbf{Step 2}: Draw $(b^{Y(\ell)}_i(q), b^{M(\ell)}_i(q))| b^{A(\ell)}_i \sim MVN(b^{(\ell)}_{t|A}(q), V^{(\ell)}_{t|A}(q))$, where
		\State \hspace{5.3em} $b^{(\ell)}_{t|A}(q)$ and $V^{(\ell)}_{t|A}(q)$ are obtained by \eqref{bcond} and \eqref{Vcond}, respectively. 
		
		\State \hspace{2.3em}\textbf{Step 3}: Update $M^{(\ell)}_{it}(q)$ 
		\State \hspace{5.3em} let $p^{(\ell)}_{it}(q) = \text{logit}^{-1}\eta^M( \mathcal{F}^{(\ell)}_{it}(q), b^{M(\ell)}_i(q); \theta^{M(\ell)})$
		\State \hspace{5.3em} draw $M^{(\ell)}_{it}(q) \sim \text{Bernoulli}(p^{(\ell)}_{it}(q))$ by setting $M^{(\ell)}_{it}(q) = \mathbbm{1} \{\psi^M_{\ell, t-h}\le p^{(\ell)}_{it}(q)\}$

        \State \hspace{2.3em}\textbf{Step 4}: Update $Y^{(\ell)}_{it}(q)$ 
		\State \hspace{5.3em} draw $Y^{(\ell)}_{it}(q) \sim f_Y(\eta^{Y(\ell)}_{it}(q), (\sigma^{(\ell)})^2)$ by 
		\State \hspace{5.3em} setting $\eta^{Y(\ell)}_{it}(q) = \eta^Y( \mathcal{F}^{(\ell)}_{it}(q), b^{Y(\ell)}_i(q); \theta^{Y(\ell)})$ and $Y^{(\ell)}_{it}(q) = \eta^{Y(\ell)}_{it}(q) + \sigma^{(\ell)} \psi^Y_{\ell, t-h}$
		
		\State \hspace{2.3em}\textbf{Step 5}: Define $$\mathcal{F}^{(\ell)}_{i,t+1}(q) = (V_i, \overline{Y}^{(\ell)}_{i,0:t} (q), \overline{M}^{(\ell)}_{i,0:t} (q), \overline{A}_{i,0:(t+1)} (q) ),$$
		\State \hspace{5.3em}  where 
		\begin{align*}
		    \overline{Y}^{(\ell)}_{i,0:t} (q) &= (\overline{Y}_{i,0:h}, \overline{Y}^{(\ell)}_{i,(h+1):t} (q))\\
		    \overline{M}^{(\ell)}_{i,0:t} (q) &= (\overline{M}_{i,0:h}, \overline{M}^{(\ell)}_{i,(h+1):t} (q))\\
		    \overline{A}_{i,0:(t+1)} (q) &= \overline{a}_{0:(t+1)}(q) 
		\end{align*}
		\State \hspace{5.3em}   and the observed equals the counterfactual during the given history, i.e. $\overline{A}_{i,0:h} = \overline{a}_{0:h}(q)$.

		\State \hspace{1em} \textbf{end for}
	\EndWhile

		\State \textbf{Step 6}:  $\{ \overline{Y}^{(\ell)}_{i,(h+1):T} (q), \overline{M}^{(\ell)}_{i,(h+1):T} (q); \ell = 1,\ldots,N_{post})\}$ are samples from the posterior predictive distribution of $(\overline{Y}_{i,(h+1):T}(q),\overline{M}_{i,(h+1):T}(q))$ under regime $q$.

	\end{algorithmic}
\end{algorithm}

\section{G-formula in Simulation}\label{simg}
\begin{align*}
\mathbb{E}[Y_{i2}(q)] =&  \iint y_2 P(Y_{i1}(q)=y_1,Y_{i2}(q)=y_2|V=v) P(V=v)dv dy_1 dy_2\\
=&  \iint y_2 P(Y_{i1}(q)=y_1,Y_{i2}(q)=y_2|V=v, b^A_i=w) P(b^A_i=w)P(V=v) dw dv dy_1 dy_2\\
=&  \iint y_2 P(Y_{i1}(q)=y_1,Y_{i2}(q)=y_2|A_{i1}=a_1, V=v, b^A_i=w) P(b^A_i=w)P(V=v) dw dv dy_1 dy_2\\
=&  \iint y_2 P(Y_{i1}=y_1,Y_{i2}(q)=y_2|A_{i1}=a_1, V=v, b^A_i=w) P(b^A_i=w)P(V=v) dw dv dy_1 dy_2\\
=&  \iint y_2 P(Y_{i2}(q)=y_2|A_{i1}=a_1,Y_{i1}=y_1, V=v, b^A_i=w)\\
& \qquad P(Y_{i1}=y_1|A_{i1}=a_1, V=v, b^A_i=w) P(b^A_i=w)P(V=v) dw dv dy_1 dy_2\\
=&  \iint y_2 P(Y_{i2}(q)=y_2|A_{i1}=a_1, A_{i2}=a_2,Y_{i1}=y_1, V=v, b^A_i=w)\\
& \qquad P(Y_{i1}=y_1|A_{i1}=a_1, V=v, b^A_i=w) P(b^A_i=w)P(V=v) dw dv dy_1 dy_2\\
=&  \iint y_2 P(Y_{i2}=y_2|A_{i1}=a_1, A_{i2}=a_2,Y_{i1}=y_1, V=v, b^A_i=w)\\
& \qquad P(Y_{i1}=y_1|A_{i1}=a_1, V=v, b^A_i=w) P(b^A_i=w)P(V=v) dw dv dy_1 dy_2\\
=& \mathbb{E}(Y_{i2}| A_{i1}=a_1, A_{i2}=a_2).
\end{align*}
\end{appendices}

\newpage
\appendix

\begin{center}
\textbf{\LARGE Supplementary Material}
\end{center}
\section{Connection with Structural Nested Models}\label{structural}

\citet{sitlani2012longitudinal} and \citet{qian2020linear} studied instantaneous treatment effect as the ``blip'' of a structural nested model (SNM), using linear mixed models as the structural model and comparing treatment paths that only differ in the treatment status at a specific time $m$, i.e. comparing $A_m = 1$ versus $A_m = 0$ in the case of binary and monotone treatment. Our proposal, on the other hand, compares the effect of treatment paths under different regimes, i.e. $\overline{A}_{0:t}$ being $\overline{a}_{0:t}(q_1)$ versus $\overline{a}_{0:t}(q_2)$, where $q_1$ and $q_2$ are the regimes of interest. The motivating application investigates the treatment effect of taking a drug continuously over time, where the causal effect is cumulative over time and thus requires a fundamentally different characterization than a structural model approach.  The instantaneous treatment effect, or the blip, can be characterized under our framework as the average causal effect comparing $q_1$ and $q_2$ where $a_t(q_1) = a_t(q_2)$ for $t\neq m$, $a_m(q_1) = 1$ and  $a_m(q_2) = 0$. Specifically, assuming $\phi_A(\overline{A}_{i,0:t}) = A_{it}$, $\phi^Y_4(\mathcal{H}_{it}) = 0$, $\tau = 1$, and a linear mixed model for a continuous outcome leads to a special case in \citet{qian2020linear} , where we will have the instantaneous subgroup treatment effect at $h+1$ conditional on information up to time $h$ being
\begin{equation}\label{special1}
    \mathbb{E}(Y_{i,h+1}|V_i,A_{i,h+1} = 1, \overline{A}_{i,0:h},\mathcal{H}_{ih}) - \mathbb{E}(Y_{i,h+1}|V_i,A_{i,h+1} = 0, \overline{A}_{i,0:h},\mathcal{H}_{ih}) = \phi^Y_2(H_{it})\beta^Y_2.
\end{equation}
Thus, the model parameter $\beta^Y_2$ has a causal interpretation marginally over the subgroup defined by $(V_i,\overline{A}_{i,0:h},\mathcal{H}_{ih})$ in this case and the MGLMM reduces to a linear structural mixed model. However, when $\phi^Y_4(\mathcal{H}_{it}) \neq 0$, equation \eqref{special1} is no longer true because $\beta^Y_2$ only remains with a causal interpretation conditional on $b^Y_{i}$, as showed in the conditional subgroup causal effect below,
\begin{align}\label{special2}
    \mathbb{E}(Y_{i,h+1}|V_i,A_{i,h+1} = 1, \overline{A}_{i,0:h},\mathcal{H}_{ih}, b^Y_i) - \mathbb{E}(Y_{i,h+1}|V_i,A_{i,h+1} = & 0, \overline{A}_{i,0:h},\mathcal{H}_{ih}, b^Y_i)\nonumber \\
    =& \phi^Y_2(H_{it})\beta^Y_2 + \phi^Y_4(H_{it})b^Y_{i1},
\end{align}
 and the conditional expectation $\mathbb{E}(b^Y_i|V_i,\overline{A}_{i,0:h},\mathcal{H}_{ih})$ is not necessarily zero.

\section{Connection with \citet{shardell2018joint}}\label{connectshardell}

\citet{shardell2018joint} demonstrated longitudinal causal inference using joint mixed-effects models, assuming shared random effects between the model components for the outcome, confounders, and treatment assignment. Their model specification is similar to the MGLMM in Section \ref{model}, i.e. with $b^A_{i0} = b^M_{i0} =  (b^Y_{i0},b^Y_{i1})$ and $\phi^M_4\equiv 0$, but  distinctively different in that $\phi^A_2$ and $\phi^M_3$ are population-level coefficients instead of observed variables. \citet{shardell2018joint} assumed sequential exchangeability conditional on the unobserved heterogeneity in the outcome progression, i.e. $(b^Y_{i0},b^Y_{i1})$, which is assumed to be proportionate to the heterogeneity in confounders and treatment assignment.  
Whereas we account for unobserved time-invariant traits in treatment assignment with the random effect $b^A_i$, assuming that it is correlated with $(b^Y_{i0},b^Y_{i1})$ and having the sequential exchangeability conditional on  $b^A_i$ instead of $(b^Y_{i0},b^Y_{i1})$. 

The assumption of no unmeasured confounders in the model of \citet{shardell2018joint} implies no treatment assignment heterogeneity. While assuming no treatment heterogeneity under MGLMM is equivalent to setting $v=0$, which is a sufficient but unnecessary condition for having no unmeasured confounders. In MGLMM, $\text{cov}(b^A_i, b^M_i) =\text{cov}(b^A_i, b^Y_i) = 0$ leads to no unmeasured confounders.  That is, even when no unmeasured confounders is true, MGLMM still allows treatment assignment heterogeneity as long as it is not correlated with the unobserved heterogeneity in biomarker dynamics $(b^Y_i, b^M_i)$; examples of such unconfounding treatment assignment heterogeneity include a patient's preference for a treatment based on personal beliefs or social stigma. On the other hand, we note that the covariances $\text{cov}(b^A_i, b^M_i)$ and $\text{cov}(b^A_i, b^Y_i)$ are estimable given $v$, the presumed variance of $b^A_i$. Henceforth, our method does partially inform the possible existence of unmeasured confounders based on the estimated covariances in the MGLMM.

When there is no treatment assignment heterogneiety, both \citet{shardell2018joint} and our method simplify to the standard g-computation of fitting only the outcome and confounders model using generalized linear mixed-effects model because the assignment mechanism becomes ignorable \citep{li2022bayesian}. Let us consider a simplified scenario of looking at the subgroup ATE at time $h+1$ conditional on history information up to time $h$, assuming no time-varying confounders and no treatment assignment heterogeneity. The subgroup ATE would not be identifiable under \citet{shardell2018joint}. The reason is as follows. Under their conditional sequential exchangeabiltiy assumption  $$Y_{h+1}(q)\perp A_{h+1} | V,\overline{A}_{0:h},\overline{Y}_{0:h}, b^Y_i,$$ we can directly identify the conditional counterfactual distribution as $$P(Y_{h+1}(q)| V,\overline{A}_{0:h},\overline{Y}_{0:h}, b^Y_i)=P(Y_{h+1}| V,\overline{A}_{0:(h+1)} = \overline{a}_{0:(h+1)}(q),\overline{Y}_{0:h}, b^Y_i).$$ However, the target quantity  represented by $P(Y_{h+1}(q)| V,\overline{A}_{0:h},\overline{Y}_{0:h})$ would not be calculable because
$$P(Y_{h+1}(q)| V,\overline{A}_{0:h},\overline{Y}_{0:h}) = \int P(Y_{h+1}(q)| V,\overline{A}_{0:h},\overline{Y}_{0:h}, b^Y_i) P(b^Y_i| V,\overline{A}_{0:h},\overline{Y}_{0:h})db^Y_i$$
and the subgroup heterogeneity distribution $P(b^Y_i| V,\overline{A}_{0:h},\overline{Y}_{0:h})$ is unknown. Whereas with our proposal, we assume a different conditional sequential exchangeabiltiy assumption  $$Y_{h+1}(q)\perp A_{h+1} | V,\overline{A}_{0:h},\overline{Y}_{0:h}, b^A_i.$$ Given the assumption of no treatment assignment heterogeneity, we know $\text{var}(b^A_i) = 0$ and consequently $\text{cov}(b^A_i, b^Y_i) = 0$, leading to $P(b^Y_i| V,\overline{A}_{0:h},\overline{Y}_{0:h}, b^A_i) = P(b^Y_i| V,\overline{A}_{0:h},\overline{Y}_{0:h}) $. As a result, the target quantity is identifiable via \eqref{gformula} as
$$P(Y_{h+1}(q)| V,\overline{A}_{0:h},\overline{Y}_{0:h}) = \int P(Y_{h+1}(q)| V,\overline{A}_{0:h},\overline{Y}_{0:h}, b^Y_i) P(b^Y_i| V,\overline{A}_{0:h},\overline{Y}_{0:h}) db^Y_i. $$

Our proposal may be viewed as an extension of Shardell and Ferrucci's \citep{shardell2018joint} work in the following aspects: (1) a softer assumption on the conditional sequential exchangeability, stratifying by $b^A_i$ instead of the random effects shared across the outcome, confounders, and treatment model, (2) model specification as MGLMM, which is more generalized and has the potential to include their joint mixed-effects model as a special case, and (3) allows the identification of subgroup causal effects when assuming no treatment assignment heterogeneity. The merits of this extension come at a price of introducing the variance of $b^A_i$ as a sensitivity parameter, and identifying subgroup causal effects under the existence of treatment assignment heterogeneity needs to be done under additional assumption on the subgroup distribution of $b^A_i$, which likely requires expert knowledge.

%Second, through using posterior probability notations, i.e. $P(\beta, G|\mathcal{D})$, show how the method integrates over both population level information and individual level trajectores. The observed data $\mathcal{D} = \cup^N_{i} \mathcal{F}_{i,n_i}$ where $n_i$ is the number of time points of subject $i$ and $N$ is the total number of subjects.

\end{document}